\documentclass[12pt]{article}
\usepackage{axodraw}
\def\header{\begin{flushleft} 
            ZU-TH \\December 2002
            \end{flushleft}}

\columnwidth\textwidth

\newcommand{\ba}{\begin{array}}
\newcommand{\ea}{\end{array}}
\newcommand{\bd}{\begin{displaymath}}
\newcommand{\ed}{\end{displaymath}}
\newcommand{\be}{\begin{equation}}
\newcommand{\ee}{\end{equation}}
\newcommand{\bea}{\begin{eqnarray}}
\newcommand{\eea}{\end{eqnarray}}

\def\qb{\bar{q}}

\def\bra{\langle}
\def\ket{\rangle}

\def\a{\alpha}
\def\b{\beta}
\def\g{\gamma}
\def\d{\delta}

\def\ve{\varepsilon}

\def\m{\mu}
\def\n{\nu}
\def\o{\omega}
\def\r{\rho}
\def\s{\sigma}
\def\t{\tau}
\def\G{\Gamma}
\def\D{\Delta}
\def\L{\Lambda}

\def\gmn{g_{\m \n}}
\def\gma{g_{\m \a}}
\def\gna{g_{\n \a}}
\def\to{\rightarrow}

\def\en{\varepsilon_{h}}
\def\es{\varepsilon_{h'}}
\def\ess{\varepsilon^{\ast}_{h''}}
\def\essm{\varepsilon^{\ast \m}_{h''}}
\def\essa{\varepsilon^{\ast \a}_{h''}}
\def\ep{\ve_\perp}
\def\N{\hat{N}}
\def\p{\hat{p}}
\begin{document}
\thispagestyle{empty}
\header \vspace*{2cm} \centerline{\Large\bf 
Consistent treatment of spin-1 mesons}
\centerline{\Large\bf 
in the light-front quark model} 
     \vspace*{2.0cm}
\centerline{\large Wolfgang Jaus}
\vspace*{0.5cm} \centerline{Institut 
f\"ur Theoretische Physik der Universit\"at Z\"urich, Winterthurerstr. 190, CH-8057 Z\"urich} 
\centerline{Switzerland}
\vspace{2cm}
\centerline{\Large\bf Abstract} \vspace*{1cm} 
\hspace{0.5cm} We analyze the matrix element of the electroweak current
between $q \qb$ vector meson states in the framework of a covariant
extension of the light-front formalism. The light-front matrix element of
a one-body current is naturally associated with zero modes, which affect
some of the form factors that are necessary to represent the Lorentz
structure of the light-front integral. The angular condition contains
information on zero modes, i.e., only if the effect of zero modes
is accounted for correctly, is it satisfied. With plausible assumptions
we derive from the angular condition several consistency conditions which can be
used quite generally to determine the zero mode contribution of form factors.
The correctness of this method is tested by the phenomenological success
of the derived form factors. We compare the predictions of our formalism 
with those of the standard light-front approach and with available data. 
As examples we discuss the magnetic moment of the
$\r$, the coupling constant $g_{D^\ast D \pi}$, and the  coupling
constants of the pseudoscalar density, $g_\pi$ and $g_K$, which provide a phenomenological 
link between constituent and current quark masses.

PACS number(s): 12.39Ki, 13.25Ft, 13.40-f, 14.65Bt

\newpage
\section{Introduction} 
\hspace{0.5cm} The light-front formalism \cite{terentev} has been used extensively
in particle and nuclear physics. It provides a conceptually simple,
phenomenological framework for the determination of the form
factors of a relativistic composite system. For practical applications
matrix elements of the current operator, which is approximated by a
one-body current, are taken between states with the
same number of constituents; these are, e.g., the confined constituent
quark and antiquark meson states, the three-quark baryon states or the
bound two-nucleon deuteron state. In general this approach does not permit
an unambiguous determination of the physical form factors, 
for it is well known \cite{karmanov1,fuda,ji1} that the 
light-front calculation of the matrix element of a one-body current 
generates a 4-vector structure that is in
general not covariant, since it contains a spurious dependence on the
orientation of the light-front. The light-front is defined in terms of the
lightlike 4-vector $\o$ by the invariant equation $\o \cdot x = 0$.
The special case $\o = (\o^0,\o^1,\o^2,\o^3) = (1,0,0,-1)$ corresponds 
to the light-front or null-plane $\o \cdot x=x^{+} = x^{0} + x^{3} = 0$.

The 4-momenta of the initial and final states are therefore not
enough to construct the Lorentz decomposition of the approximate
matrix element, the unphysical 4-vector $ \o$ is required too. This
construction does not only lead to additional, unphysical form factors,
but generates also a spurious dependence  on $\o$ of some of the form 
factors.
This problem is closely associated with the violation of rotational
invariance in the computation of the matrix element
of a one-body current \cite{leutwyler,frankfurt}.

However, full Lorentz covariance could be restored even at this level of 
approximation if it would be possible to include the effect of the 
associated zero modes \cite{yan}. In Ref. \cite{jaus1} we have investigated the
relationship between a simple (but unrealistic) covariant model for
$q \qb$ mesons and its light-front representation in 1-loop order.
In a frame where the momentum transfer $q^+ =q^0 + q^3 =0$, the model
permits the calculation of the $\d (x)$ contributions, where 
$p'^+_1 = x  P'^+$ is the plus component of the constituent quark
momentum and $P'^+$ is the plus component the momentum of the initial meson state. The
$\d (x)$ or zero mode terms are naturally associated with the light-front
matrix element of a one-body current and restore full covariance. They
minimally account for neglected interaction effects which is obvious from the work
in Refs. \cite{melo,brodsky,ji2} where the zero mode contributions
proportional to $\d (x)$ have been interpreted as residues of virtual
pair creation processes in the limit $q^+ = 0$.
While these model calculations \cite{jaus1,melo,brodsky,ji2}
have been used to investigate the role of zero modes in a light-front approach, they did not provide
a practical method for a realistic phenomenology. It is the purpose of the present work to develop
a method that permits the calculation of the contribution of the zero modes associated with
the matrix element of a one-body current and, consequently, a consistent determination
of the physical form factors.

Note that the standard light-front (SLF) formalism uses only the matrix elements of the plus
component of the one-body current and requires all constituents to be on their respective
mass shells. These rules ignore the effect of zero modes and the covariance of the matrix
element is lost. Nevertheless,
the zero mode contributions can be avoided for transitions that involve only
pseudoscalar mesons if the hadronic form factors are calculated from
the plus components of the respective currents, but this is no longer
true for composite spin-1 systems. An impressive example is provided
by three different light-front calculations of the elastic electromagnetic
form factors of the deuteron in Refs. \cite{GK,CCKP,BH}. The starting point
in each case is the plus component of the matrix element of the one-body
electromagnetic current, nevertheless the final results are different. Such
an outcome is possible since the number of independent helicity amplitudes
is larger than the number of physical form factors. However, the requirement 
of covariance imposes a nontrivial dynamic constraint on the helicity amplitudes
which is known as the angular condition \cite {GK,CCKP}. If the angular condition
is fulfilled, different prescriptions for the calculation of the form factors
lead to the same result. The standard light-front approach in general violates
the angular condition and thus signals again that it is not covariant.

An interesting analysis of the spin-1 system has recently been published by
Bakker, Choi and Ji \cite{ji3,ji4}. In Ref. \cite{ji3} the frame dependence of the
angular condition is investigated. In Ref. \cite{ji4} spin-1 form factors are
analyzed with respect to different prescriptions, different choices of polarization
vectors and reference frames. In \cite{ji4} a covariant model is used, which (as
far as the triangle diagram is concerned) is similar to ours \cite{jaus1}, to guide
the corresponding light-front calculation. A particular prescription is found to 
derive the physical form factors, that is not affected by zero mode contributions 
and coincides exactly with the result of the covariant model calculation.

The derivation \cite{ji4} of this outcome is model dependent and we 
shall show in the present paper that it is of general validity only for the simple
$q \qb$-vector meson vertex chosen in Ref. \cite{ji4}, but if the electromagnetic 
form factors of a vector meson, $F_1(q^2)$, $F_2(q^2)$ and $F_3(q^2)$, are calculated 
with the standard light-front $q \qb$-vector meson vertex, the magnetic dipole form factor 
$F_2(q^2)$  requires zero mode contributions. Our analysis is
based on the formalism which we have developed in Ref. \cite{jaus1}. At first
we shall investigate electroweak transitions between $q \qb$ mesons of spin 1.
The angular condition is shown to lead to several constraints on those parts of
the matrix element of the one-body current that depend on $\o$, and which must be completed
by an appropriate account of zero modes. In this manner spurious contributions to
form factors can be eliminated completely. The angular condition is of crucial
importance for our analysis: it not only checks the covariance of a procedure, but
it contains information on the effect of zero modes of which we shall take
full advantage. This method permits a consistent
light-front calculation of the transition form factors of vector
mesons in the 1-loop approximation.

There is an alternative and quite different scheme to deal with the problem
of the violation of covariance of the SLF formalism, which has been reviewed
in Ref. \cite{karmanov1} (we shall refer to it by the label CDKM).
In order to treat the complete Lorentz structure of a hadronic matrix
element the authors of Ref. \cite{karmanov1} have developed a method to identify and separate
spurious contributions and to determine the physical, i.e. $\o$-independent
contributions to the hadronic form factors and coupling constants. We shall
compare the results of our work with those of the CDKM approach.

In \cite{jaus1} we have found that in the determination of form factors
one has to deal with two classes. There is one class of form factors like
the charge form factor of a pseudoscalar meson, $V(q^2)$ and $A_2(q^2)$ for transitions between vector
and pseudoscalar mesons,
and the pseudoscalar coupling constant $f_P$ that are free of zero mode contributions. 
These form factors are predicted unambiguously
in 1-loop order by the standard light-front approach, and every covariant extension of
the SLF formalism (like the approach of this work or that of CDKM) must reproduce these results.
Another class of form factors like $F_2(q^2)$, $A_1(q^2)$ and the vector coupling constant $f_V$
are associated with zero modes. For their reliable prediction in 1-loop order the
corresponding zero mode contribution must be known. 
It is somewhat surprising that the angular condition,
which has been established for transitions between vector mesons, can be used
not only to determine the zero mode contributions to the magnetic form factor
of a vector meson $F_2(q^2)$, but also those to
$f_V$ and  $A_1(q^2)$.
The form factors that we shall determine in such a manner are different from
those obtained in the SLF or CDKM schemes, both of which ignore zero mode contributions.

In Sec.II we present a brief summary of the basic formalism for the treatment of the
transition form factors for spin-1 mesons. We discuss the angular condition, and with
plausible assumptions we derive from it several consistency conditions which can be
used quite generally 
to determine effective zero mode contributions. The formal
machinery required for the Lorentz decomposition of the light-front matrix element
has been collected in the Appendices A and B. In Sec.III we use this method to calculate
the zero mode contributions  associated with $f_V$ and  $A_1(q^2)$, and derive the formulas for
these quantities in our formalism, and also
in the SLF and CDKM schemes. In Sec.IV we take the magnetic moment of
the rho meson and the coupling constants $g_{VP\pi}$ as examples to compare quantitative
predictions of our approach with those of the SLF and CDKM schemes.
We also briefly discuss the coupling constant of the pseudoscalar density,
$g_P$, whose zero mode contribution
can be determined by our method, and which provides a phenomenological link between
constituent and current quark masses. 
We conclude this work in Sec.V with a summary of our analysis.

\section{Transition form factors for spin-1 mesons}
\setcounter{equation}{0}

\subsection{Angular condition for electroweak transitions}

\hspace{0.5cm} We shall generalize the procedure of \cite{ji4} for the
matrix element of an electroweak current $V^\m=\qb'' \g^\m q'$ between initial and final
$q \qb$ vector meson states of 4-momentum, mass and helicity $(P',M',h')$
and $(P'',M'',h'')$, respectively. Its most general form is represented in
terms of appropriate transition form factors as

\bea
G^\m_{h'' h'} &=& \bra P'',h''| V^\m |P',h'\ket \nonumber\\ 
&=& \ess \cdot \es \: P^\m F_1 (q^2)
                                + (\es^\m \: \ess \cdot P +\essm \: \es \cdot P ) F_2 (q^2) \nonumber\\
& & +\frac{1}{2 M' M''} (\ess \cdot P)(\es \cdot P) P^\m F_3 (q^2) \nonumber\\
& &                                + (\es^\m \: \ess \cdot P -\essm \: \es \cdot P ) F_4 (q^2) \nonumber \\
& & + \ess \cdot \es \: q^\m W_1 (q^2) 
                                +\frac{1}{2 M' M''} (\ess \cdot P)(\es \cdot P) q^\m W_2 (q^2)\, , \nonumber\\ 
\eea
where $P=P' +P''$ and $q= P' -P''$. For the special case of an electromagnetic
transition $F_4 (q^2) = W_i (q^2) = 0$ for $i=1,2$.

We shall always impose the condition $q^+ =0$, which means that the form factors are known
only for spacelike momentum transfer $q^2 = -q^2_\perp \le 0$. In Ref. \cite{jaus2} we have
proposed to rewrite a form factor as an explicit function of $q^2$ and analytically continue
from timelike to spacelike momentum transfer, and verified this procedure in a covariant model
calculation in Ref. \cite{jaus1}. 

The 4-momentum $P'$ of a meson of mass $M'$ in terms of
light-front components is $P' = (P'^- ,P'^+ ,P'_\perp )$, where $P'^\pm = P'^0 \pm P'^3$ 
and $P'^2 = P'^+ P'^- - P'^2_\perp  = M'^2$. 
The scalar product of two 4-vectors A and B is given by $A \cdot B= \frac{1}{2}A^+ B^-
+\frac{1}{2}A^- B^+ -A_\perp B_\perp$.
We use a reference frame, where
$q_\perp = (q^1 ,q^2 ) = (Q,0)$, $P'_\perp =0$ and $P''_\perp = -q_\perp$.
The light-front polarization vector depends upon the momentum of the vector meson. For
example, $\es \equiv \ve (P',h') = (\ve ^- ,\ve ^+ ,\ve_\perp )$ is given by

\bea
\ve (P',\pm 1) &=& \Big( \frac{2}{P'^+} \ep (\pm 1) P'_\perp,0,\ep (\pm 1) \Big) \\
\ep (\pm 1) &=&  \frac{\mp 1}{\sqrt{2}} (1,\pm i) \nonumber \\
\ve (P', 0) &=& \frac{1}{M'}  \Big( \frac{-M'^2 +P'^2_\perp}{P'^+},P'^+,P'_\perp \Big) ,
\eea
and an analogous expression holds for $\ess$.

In this subsection we shall work only with the plus component of the current, and the
corresponding matrix element is defined by $G^+_{h'' h'} = \o \cdot G_{h'' h'}$. Since
$\o \cdot q = q^+ =0$, there are five independent helicity amplitudes which are related
to the four form factors, which can be determined in this manner, as
\bea
G^+_{++} &=& -2M' \Big( F_1 + \frac{Q^2}{2}\frac{F_3}{2M'M''} \Big) \, , \\
G^+_{+-} &=& M'Q^2 \frac{F_3}{2M'M''} \, ,\\
G^+_{+0} &=& \sqrt{2} Q F_1 +\frac{Q}{\sqrt{2}} (F_2 -F_4 )+\frac{Q}{\sqrt{2}}(Q^2
             -q \cdot P ) \frac{F_3}{2M'M''} \, , \\
G^+_{0+} &=& -\frac{\sqrt{2} M' Q}{M''} F_1 -\frac{M' Q}{\sqrt{2} M''} (F_2 +F_4)
             -\frac{M' Q}{\sqrt{2} M''}(Q^2 + q \cdot P)  \frac{F_3}{2M'M''} \, , \\
G^+_{00} &=& -\frac{1}{M''} (M'^2 +M''^2 -Q^2) F_1 +\frac{Q^2}{M''} F_2 
              +\frac{q \cdot P}{M''} F_4  \nonumber\\
& & +\frac{1}{2M''} (Q^4 - (q \cdot P)^2)\frac{F_3}{2M'M''} \, .
\eea
These equations can be solved uniquely for the four form factors only if the helicity
amplitudes obey the angular condition
\bea
\D(q^2) & \equiv & G^+_{00}-\frac{Q^2 -q\cdot P}{\sqrt{2} M'' Q}G^+_{+0}
         +\frac{Q^2 +q\cdot P}{\sqrt{2} M' Q}G^+_{0+} \nonumber\\
& &      -\frac{M'^2 +M''^2 +Q^2}{2 M' M''}G^+_{++}
         -\Big( \frac{M'^2 +M''^2}{2M' M''}-\frac{3(q \cdot P)^2}{2M'M''Q^2} \Big) G^+_{+-}
\nonumber\\
&=& 0 \, ,
\eea
where $Q^2 = -q^2$.
For $q \cdot P = M'^2 -M''^2 =0$ and $G^+_{0+}=-G^+_{+0}$, Eq.(2.9) turns 
into the usual angular condition
\cite{GK,ji4} for electromagnetic transitions.

\subsection{Light-front matrix elements of the electroweak current}

\hspace{0.5cm} We shall use the same notation as in Ref. \cite{jaus1}. The  $q\qb$ meson
of mass $M'$ and 4-momentum $P'$ is composed of off-shell constituent quarks with masses $m'_1,m_2$
and 4-momenta $p'_1,p_2$, respectively, with $p'_1 +p_2 =P'$. The appropriate variables for
the internal motion of the constituents, $(x,p'_\perp)$, are defined by
\bea
p'^+_1  &=& xP'^+ \quad , \quad p_2 ^+ = (1-x)P'^+ \, ,\nonumber\\
p'_{1 \perp} &=& xP'_\perp + p'_\perp \quad , \quad p_{2 \perp} = (1-x)P'_\perp - p'_\perp \, , \nonumber
\eea
and the kinematic invariant mass is
\[
M'^2_0  =  \frac{p'^2_\perp  + m'^2_1  }{x}
                         + \frac{p'^2_\perp  + m_2 ^2 }{1-x} \, . 
\]


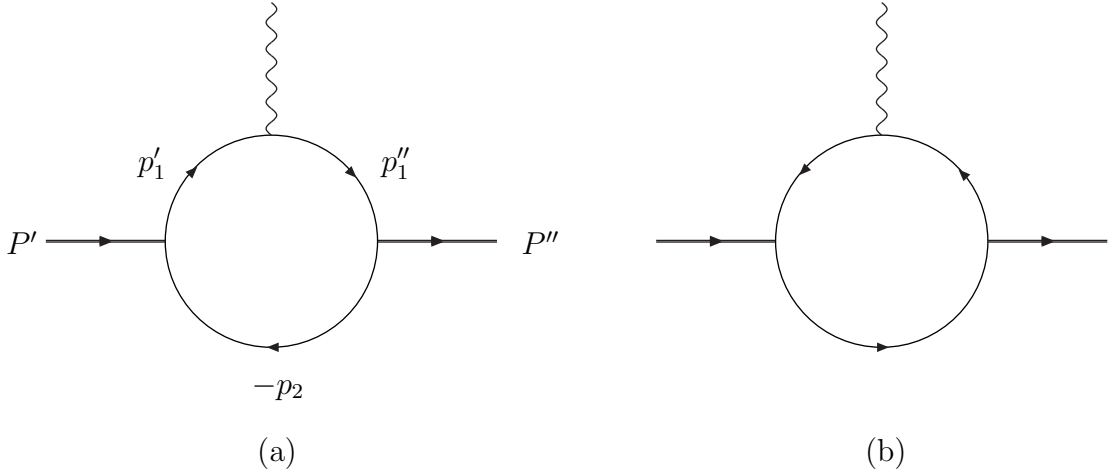
\begin{figure}[h] 
\begin{center}
\begin{picture}(400,180)



\Text(0,90)[l]{$P'$}
\ArrowLine(15,90.3)(60,90.3)
\ArrowLine(15,89.7)(60,89.7)
\ArrowLine(140,90.3)(185,90.3)
\ArrowLine(140,89.7)(185,89.7)
\Text(195,90)[l]{$P''$}

\GCirc(100,90){40}{1}

\ArrowLine (101,50)(100,50)
\ArrowLine (70,116)(71,117)
\ArrowLine (130,116)(131,115)

\Text(50,120)[l]{$p'_1$}
\Text(142,120)[l]{$p''_1$}
\Text(93,35)[l]{$-p_2$}


\Photon(100,130)(100,180){2}{5}


\Text(95,10)[l]{(a)}


\ArrowLine(245,90.3)(290,90.3)
\ArrowLine(245,89.7)(290,89.7)
\GCirc(330,90){40}{1}
\ArrowLine(370,90.3)(415,90.3)
\ArrowLine(370,89.7)(415,89.7)

\ArrowLine (330,50)(331,50)
\ArrowLine (301,117)(300,116)
\ArrowLine (361,115)(360,116)


\Photon(330,130)(330,180){2}{5}

\Text(325,10)[l]{(b)}

\end{picture}
\caption{The one-loop contributions to the transition between $q\qb$ mesons .}
\end{center} 
\end{figure}


The matrix element (2.1) is given in the 1-loop approximation, corresponding to the
diagram of Fig.1a, as a light-front integral, which we shall denote by $\hat G^\m_{h'' h'}$.
For the transition between an initial vector meson with internal variables 
and masses of its constituent quarks $(x,p'_\perp ,m'_1 ,m_2 )$ and a final vector
meson with the corresponding quantities $(x,p''_\perp ,m''_1 ,m_2 )$ the momentum integral
$\hat G^\m_{h'' h'}$ is given by \cite {jaus1}
\be
\hat G^\m_{h'' h'} = \frac{N_c}{16\pi^3}\int^1_0 dx \int d^2 p'_\perp
\frac{\es^\b tr \left\{ \G'_\b(-\not p_2+m_2) \G''_\a
(\not p''_1 +m''_1) \g^\m (\not p'_1 +m'_1) \right\}
\essa}{(1-x) \N'_1 \N''_1} \, ,
\ee
and $N_c$ denotes the number of colors. For electromagnetic transitions there is an analoguous
contribution due to the diagram of Fig.1b.

Eq.(2.10) is computed at the pole of the spectator quark
and we shall use 
$\p'_1,\p''_1$ and $\p_2$ to denote the restricted 4-vectors
\be 
\p_2 =(\frac{m^2_{2\perp}}{p^+_2},p^+_2,p_{2\perp}) \quad , \quad
\p'_1 =P'-\p_2 \quad , \quad
\p''_1 =\p'_1 -q \, , 
\ee
where $m^2_{2\perp}=p^2_{2\perp}+m^2_2$. We shall use also the abbreviations
$m'^2_{1\perp}=p'^2_{1\perp}+m'^2_1$ and $m''^2_{1\perp}=p''^2_{1\perp}+m''^2_1$.
In the numerator of the integrand of Eq.(2.10) we
have kept the unrestricted constituent momenta for reasons that will be explained below.
It follows from Eq. (2.11) that
\bea
\N_2 &=& \p^2_2 -m^2_2 =0 \, , \nonumber \\
\N'_1 &=& \p'^2_1 -m'^2_1 =x(M'^2-M'^2_0 ) \, , \\
\N''_1&=& \p''^2_1-m''^2_1=x(M''^2-M''^2_0) \, . \nonumber
\eea
The vector vertex operators for $^3S_1$-state mesons are given in Ref. \cite{jaus1} as
\bea
\G'_\b \es^\b &=& -h'_0
\left\{ \g_\b - \frac{1}{D'} (p'_1 -p_2)_\b \right\} \es^\b \, , \nonumber\\ 
\G''_\a \essa &=& -h''_0
\left\{ \g_\a - \frac{1}{D''} (p''_1 -p_2)_\a \right\} \essa \, , 
\eea
where
\bea
h'_0 &=& \left[ \frac{M'^4_0 -(m'^2_1 -m^2_2 )^2}{4 M'^3_0} \right]^{1/2}
\frac{M'^2-M'^2_0}{[M'^2_0 -(m'_1 -m_2 )^2 ]^{1/2}} \phi (M'^2_0 ) \, , \\
D' &=& M'_0 +m'_1 +m_2 \, ,
\eea
and similar equations for $h''_0$ and $D''$. The orbital wave function is assumed
to be a simple function of the kinematic invariant mass as
\be
\phi (M'^2_0 ) = Z' exp(-M'^2_0 /(8 \b'^2) ) \, ,
\ee
where $Z'$ is the normalization constant (see Eq.(4.2)) and the parameter $1/\b'$
determines the confinement scale.

The helicity amplitude can now be rewritten as
\bea
\hat{G}^\m_{h'' h'} &=&\frac{N_c}{16\pi^3} \int^1_0 dx \int d^2 p'_\perp \frac{h'_0h''_0}
{(1-x) \N'_1 \N''_1} \nonumber\\
& &  \Big( A^\m -\frac{1}{D'}B^\m -\frac{1}{D''}C^\m +\frac{1}{D' D''}D^\m \Big) \, ,
\eea
where
\bea
A_\m&=& \es^\b \essa tr\left[\g_\a (\not p''_1+m''_1)\g_\m(\not p'_1+m'_1)\g_\b (-\not p_2+m_2)
\right] \, , \nonumber\\
B_\m&=& \es \cdot (p'_1 -p_2)\; \essa \; tr\left[\g_\a (\not p''_1+m''_1)\g_\m(\not p'_1+m'_1) 
(-\not p_2+m_2)
\right] \, , \nonumber\\
C_\m&=& \ess \cdot (p''_1 -p_2) \; \es^\b \; tr\left[(\not p''_1+m''_1)\g_\m(\not p'_1+m'_1)\g_\b 
(-\not p_2+m_2)
\right] \, , \nonumber\\
D_\m&=& \es \cdot (p'_1 -p_2) \; \ess \cdot (p''_1 -p_2) \; 
tr\left[ (\not p''_1+m''_1)\g_\m(\not p'_1+m'_1) (-\not p_2+m_2)
\right] \, . \nonumber\\
\eea
With $p_2 =P'-p'_1$ and $p''_1 =p'_1 -q$ the traces of Eq.(2.18) can be expressed in terms
of the tensors $p'_{1 \m}$, $p'_{1 \m}p'_{1 \n}$ and $p'_{1 \m}p'_{1 \n}p'_{1 \a}$, whose
decompositions into the 4-vectors $P$, $q$ and $\o$ are given in Appendix A. Special care is
taken to trace the $p'^-_1$ dependence, which is encoded in the functions $B^{(m)}_n$ and $C^{(m)}_n$.
We emphasize that only the decomposition coefficients which are combined with the 4-vector $\o$,
namely $B^{(m)}_n$ and $C^{(m)}_n$, depend on $p'^-_1$ and behave like $(p'^-_1)^i (p'^+_1)^j$.
For the $B$-functions $i \le j$, which means that there is no zero mode contribution and the
value of $B^{(m)}_n$ can be calculated unambiguously at the spectator quark pole, Eq.(2.11).  
For the C-functions $i \ge j+1$, and the value of $C^{(m)}_n$ is the sum of a spectator quark 
pole term and an unknown zero mode contribution, 
as we have shown in a model calculation in Ref. \cite{jaus1}.

Note that if the integral (2.17)
is calculated at the pole of the spectator quark it is known exactly: the minus components
of the internal momenta are
given by Eq.(2.11), in particular $p'^-_1 =\p'^-_1 =P'^- -m^2_{2\perp}/p^+_2$, and the
$C$-functions can be evaluated in this case. However, such a prescription misses entirely the
effect of zero modes and violates rotational invariance, i.e., is not Lorentz covariant.
Full covariance is restored by the zero mode terms of the $C$-functions.

Another prescription to fix the minus components of the constituents is used in the
standard light-front formalism (and the CDKM scheme), where all quarks are put on their
respective mass shells, i.e., $p'^-_1=m'^2_{1\perp}/p'^+_1$, $p''^-_1=m''^2_{1\perp}/p''^+_1$
and $p^-_2=m^2_{2\perp}/p^+_2$. This is the prescription used most often in the literature.
The resulting light-front integral is known exactly, but it
contains spurious contributions due to unresolved zero mode effects.

In our approach we shall not fix the value of $p'^-_1$ and shall use the $C$-functions as a 
convenient parametrization for the contribution of zero modes.
But it is not known as yet how to calculate the contribution of the zero modes associated with 
a light-front integral like (2.17). While various model calculations \cite{jaus1,melo,brodsky,ji2}
have been used to investigate the role of zero modes in a light-front approach, they did not provide
a practical method for a realistic phenomenology. We shall address this problem in the present work.

Consequently, the light-front integral (2.17) must be represented also in terms of the 4-vectors
$P$, $q$ and $\o$ :
\bea
\hat{G}^\m_{h'' h'} 
&=& \ess \cdot \es \: P^\m F_1 (q^2)
                                + (\es^\m \: \ess \cdot P +\essm \: \es \cdot P ) F_2 (q^2) \nonumber\\
& & +\frac{1}{2 M' M''} (\ess \cdot P)(\es \cdot P) P^\m F_3 (q^2) \nonumber\\
& &                                + (\es^\m \: \ess \cdot P -\essm \: \es \cdot P ) F_4 (q^2) \nonumber \\
& & + \frac{1}{\o \cdot P} (\es^\m \: \ess \cdot \o +\essm \: \es \cdot \o ) H_1 (q^2) \nonumber\\
& & +  \frac{1}{\o \cdot P} (\es^\m \: \ess \cdot \o -\essm \: \es \cdot \o ) H_2 (q^2) \nonumber \\
& & + q^\m (\cdots) + \o^\m (\cdots) \, , 
\eea
where we have omitted all terms proportional to $q^\m$ and $\o^\m$, and we have used the same 
notation for the form factors $F_i$ regardless if they represent the approximation (2.17) or 
the exact helicity amplitude (2.1). 
The approximate helicity amplitude $\hat{G}^\m_{h'' h'}$ consists of an $\o$ independent part, and
a part that does depend on $\o$ and whose construction requires additional, unphysical form
factors. However, this separation is not unique. 
For example, in writing down Eq.(2.19) we have used the identity
\bea
\frac{1}{\o \cdot P} P^\m \; \en \cdot \o &=& \en^\m - \frac{q^\m}{q^2} \Big( \en \cdot q 
                                              -q \cdot P \; \frac{\en \cdot \o}{\o \cdot P} \Big) \nonumber\\
& &-\frac{\o^\m}{\o \cdot P} \Big(\en \cdot P - \en \cdot q \; \frac{q \cdot P}{q^2}
 -\en \cdot \o \; \frac{P^2}{\o \cdot P} \nonumber\\ 
& & +\en \cdot \o \; \frac{(q\cdot P)^2}{q^2 \; \o \cdot P} \Big)\nonumber\\
& &-\frac{1}{q^2 (\o \cdot P)^2} \; \ve^{\m \a \b \n}\o_\a q_\b P_\n \; \ve_{\r \t \s \g}\en^\r \o^\t q^\s P^\g \, ,
\eea
in order to absorb the contributions due to the terms proportional to
\bea
\frac{1}{\o \cdot P} \; P_\m \; \es \cdot \o \; \ess \cdot P \; , \; \frac{1}{\o \cdot P}\; P_\m \; 
\ess \cdot \o \; \es \cdot P \;  , \;
\frac{1}{(\o \cdot P)^2}\; P_\m \; \es \cdot \o \; \ess \cdot \o \; , \nonumber\\ 
\eea
into the form factors $F_2$, $F_4$, $H_1$ and $H_2$. 

We have made no reference to the $\ve$-tensor
terms in Eq.(2.19) and shall treat them as unphysical contributions which can be neglected.
This procedure is well justified for the longitutinal mode since the  $\ve$-tensor
term vanishes for $h=0$, because
\be
\frac{1}{\o \cdot P} \; \ve_{\r \t \s \g}\en^\r \o^\t q^\s P^\g = i h \; \en \cdot q \, .
\ee

For the transverse mode the decomposition into physical and unphysical contributions, resulting
from the identity (2.20), is ambiguous, but is consistent with the result obtained for the
longitutinal mode. In particular, the relationship between the four form factors $F_i(q^2)$
and the plus components of the helicity amplitudes $\hat{G}^+_{h'' h'}$ agree exactly with Eqs.(2.4)-(2.7),
which means that
\[
\hat{G}^+_{++}=G^+_{++}\; , \; \hat{G}^+_{+-}=G^+_{+-}\; , \; \hat{G}^+_{+0}=G^+_{+0} \; , \;
\hat{G}^+_{0+}=G^+_{0+} \; ,
\]
where the difference between the exact and the approximate helicity amplitude is understood.
Only the amplitude $\hat{G}^+_{00}$ depends on one of the unphysical form factors in the
following way:
\[
\hat{G}^+_{00} = G^+_{00} + \frac{1}{M''} H_1 \, .
\]
It is therefore not surprising that the angular condition (2.9) requires the unphysical 
form factor $H_1$ to vanish:
\be
\D (q^2) = \frac{1}{M''} \; H_1(q^2) = 0 \, .
\ee

The determination of the light-front integrals for
the form factors $F_i(q^2)$ is straightforward; it can be derived
from the detailed decomposition of the traces (2.18), which is exhibited for the 
special case of an electromagnetic transition in Appendix B. 
Note that the expressions for $F_2$ 
(and $F_4$) contain $C$-functions, which means that both form factors
are affected by zero modes.

Only the amplitude $\hat{G}^+_{00}$ contains admixtures of unphysical form factors,
which confirms the result of Ref. \cite{ji4} that the light-front analysis that uses the plus
components of the helicity amplitudes, avoiding the $(h'',h')=(0,0)$ component, is preferred
for model calculations. The form factors $F_i$ derived according to this prescription
agree exactly with our results. However, the zero mode problem cannot be avoided since, as
noted above, the calculation that uses the vertex operators (2.13) leads to form factors
that require zero mode contributions (these are related to the second term of the vertex
operator).

In the CDKM scheme of Refs. \cite{karmanov1,melikov1} the helicity amplitude is represented
also in terms of the 4-vectors $P$, $q$ and $\o$, and
the $\o$-dependent contributions are separated and omitted entirely. 
But this approach has the same problems:
a unique separation is not possible, and the form factor $F_2$ is in general affected by
zero modes.

We shall now discuss the form factor $H_1(q^2)$ in greater detail. It
can be represented in terms of the functions $B^{(m)}_n$ and
$C^{(m)}_n$, which have been defined in Appendix A. According to the representation used in Eq.(2.17)
one finds
\be
H_1(q^2) = H^A_1 +H^B_1 +H^C_1 +H^D_1 \, ,
\ee
with
\bea
H^A_1(q^2) &=& \frac{N_c}{16\pi^3} \int^1_0 dx \int d^2 p'_\perp \frac{h'_0h''_0}
{(1-x) \N'_1 \N''_1} 
(-4) \Big( Y - m'_1 m''_1 C^{(1)}_1 \Big) \, ,\nonumber\\
\eea
where
\be
Y=2C^{(3)}_1 +q^2 (C^{(3)}_2 -C^{(2)}_1 )+P^2 \; B^{(3)}_1 +2q \cdot P \; B^{(3)}_2 
-q \cdot P \; B^{(2)}_1 \, .
\ee
For $H^B_1$, $H^C_1$ and $H^D_1$  one finds
\bea
H^B_1(q^2) &=& \frac{N_c}{16\pi^3} \int^1_0 dx \int d^2 p'_\perp \frac{h'_0h''_0}
{(1-x) \N'_1 \N''_1} \nonumber\\
& & \times \frac{-2}{D'} \Big\{-2(m'_1 +m_2) \Big[ Y - m'_1 m''_1 C^{(1)}_1 \Big] +2m''_1 C^{(1)}_1 N'_1 \nonumber\\
& &~~~~~~~~+(m'_1 -m''_1) \Big[ (q \cdot P +q^2)C^{(2)}_1 +(P^2 +q \cdot P)B^{(2)}_1 \Big] \nonumber\\
& & -(q^2 +q \cdot P)m'_1 C^{(1)}_1 \Big\} \, , 
\eea
\bea
H^C_1(q^2) &=& \frac{N_c}{16\pi^3} \int^1_0 dx \int d^2 p'_\perp \frac{h'_0h''_0}
{(1-x) \N'_1 \N''_1} \nonumber\\
& & \times \frac{-2}{D''} \Big\{-2(m''_1 +m_2) \Big[ Y - m'_1 m''_1 C^{(1)}_1 \Big] +2m'_1 C^{(1)}_1 N''_1 \nonumber\\
& &~~~~~~~~-(m'_1 -m''_1) \Big[ (q \cdot P -q^2)C^{(2)}_1 +(P^2 -q \cdot P)B^{(2)}_1 \Big] \nonumber\\
& & -(q^2 -q \cdot P)m'_1 C^{(1)}_1 \Big\} \, , 
\eea
\bea
H^D_1(q^2) &=& \frac{N_c}{16\pi^3} \int^1_0 dx \int d^2 p'_\perp \frac{h'_0h''_0}
{(1-x) \N'_1 \N''_1} \nonumber\\
& & \times \frac{4}{D' D''} \Big\{ \Big[ q^2-M'^2-M''^2+2N_2-(m'_1-m''_1)^2+(m'_1+m_2)^2 \nonumber\\
& &~~~~~~~~~~~~+(m''_1+m_2)^2 \Big] (2C^{(3)}_1 +C^{(3)}_3) \nonumber\\
& &~~~~~~~~~~~~- \frac{1}{2}\Big[ q^2-N'_1-N''_1-(m'_1-m''_1)^2 \Big] C^{(2)}_2 \Big\} \, .
\eea

\subsection{Determination of effective zero mode contributions }

\hspace{0.5cm} Obviously, the angular condition (2.23) constrains a combination of
$C$-functions and consequently contains information on the effect of
zero modes. In order to exploit this hint in more detail, we use the fact 
(compare for example Refs. \cite{jaus3,jaus1}) that the light-front integral (2.10)
can be derived, in principle, from an explicitly covariant, 4-dimensional momentum
integral: The latter is expressed in terms of light-front variables and carried out
by contour methods in the complex $p'^-_1$ plane. Closing the contour in the upper
$p'^-_1$ plane (it is assumed that the vertex functions $h'_0$ and $h''_0$
have no poles there) ensures that the momentum integral is given by the residue of
the spectator quark pole, corresponding to putting quark 2 on the mass shell.
The resulting residue coincides with the light-front integral (2.10) with the prescription (2.11).
If the light-front integral could be completed by the associated zero mode contributions,
the result would be equivalent to the original, covariant 1-loop integral.
We have argued that the zero modes are related to the $p'^-_1$ dependent terms of the
traces in the representation (2.17) of the light-front integral, and we have encoded
this $p'^-_1$ dependence in terms of $C$-functions. The requirement of covariance then
leads to the angular condition (2.23).

However, each of the four trace terms $A_\m$, $B_\m$, $C_\m$ and $D_\m$ corresponds
separately to a covariant, 4-dimensional structure in the sense explained above. Each individual
contribution to the light-front integral (2.17) must be consistent with the requirement of 
covariance, which immediately leads to the conditions
\be
H^A_1 =H^B_1 =H^C_1 =H^D_1 =0 \, ,
\ee
which, of course, are consistent with the angular condition (2.23).

We can go further and use also the facts, that the
zero mode contributions do not depend on the details of the spinor structure of the vertex operators,
and that different combinations of vertex spin structures may generate the same formal decomposition 
(2.19) of the light-front integral. For example, we can formally replace the 
trace $A_\m \equiv A_\m(VVV)$, Eq.(2.18), by
three different expressions:
\bea
A_\m(AVA)&=& \es^\b \essa tr\left[\g_\a \g_5 (\not p''_1+m''_1)\g_\m(\not p'_1+m'_1)\g_\b \g_5 
(-\not p_2+m_2) \right] \nonumber\\
&=& A_\m(VVV) |_{m_2 \to -m_2} \, , \\
A_\m(VAA)&=& \es^\b \essa tr\left[\g_\a \g_5 (\not p''_1+m''_1)\g_\m \g_5 (\not p'_1+m'_1)\g_\b  
(-\not p_2+m_2) \right] \nonumber\\
&=& A_\m(VVV) |_{m'_1 \to -m'_1} \, , \\
A_\m(AAV)&=& \es^\b \essa tr\left[\g_\a  (\not p''_1+m''_1)\g_\m \g_5 (\not p'_1+m'_1)\g_\b \g_5 
(-\not p_2+m_2) \right] \nonumber\\
&=& A_\m(VVV) |_{m''_1 \to -m''_1} \,. 
\eea
Obviously, the traces $A_\m(AVA)$, $A_\m(VAA)$ and $A_\m(AAV)$ are obtained from \\
$A_\m(VVV)$ by 
changing the
signs of $m_2$, $m'_1$ and $m''_1$, respectively, while the $p'^-_1$ dependent terms remain
unchanged. 
There are corresponding consistency conditions
that can be derived from $H^A_1 \equiv H^A_1(VVV) =0$, Eq.(2.30), by the analogous change of the sign
of one of the quark masses. In this manner one finds that
\be
H^A_1(AVA) =H^A_1(VAA) =H^A_1(VAA) =0 \, .
\ee

The same procedure can be applied to the traces $B_\m$, $C_\m$ and $D_\m$ with analoguous results
(modulo an irrelevant overall change of sign). Consequently, the consistency conditions (2.30) are
invariant under the change of sign of one of the quark masses in that part of the light-front integral
which depends on one of the traces (2.18).

Since the functions $B^{(m)}_n$ and $C^{(m)}_n$ depend only on the squared quark masses, this sign 
invariance of the consistency condition $H^A_1 =0$ means, that the two terms of Eq.( 2.25) must vanish separately:
\bea
\int^1_0 dx \int d^2 p'_\perp \frac{h'_0h''_0}{(1-x)\N'_1 \N''_1} \; Y &=& 0 \, , \\
\int^1_0 dx \int d^2 p'_\perp \frac{h'_0h''_0}{(1-x)\N'_1 \N''_1} \; C^{(1)}_1 &=& 0 \, .
\eea
Similarly, the sign invariance of the consistency conditions $H^B_1 =0$ and $H^C_1 =0$, Eqs.(2.27), (2.28),
leads to the relations:
\bea
&& \int^1_0 dx \int d^2 p'_\perp \frac{h'_0h''_0}{(1-x)\N'_1 \N''_1}\frac{1}{D'} Y \nonumber\\ 
&&= \int^1_0 dx \int d^2 p'_\perp \frac{h'_0h''_0}{(1-x)\N'_1 \N''_1}
\frac{1}{D''} Y =0 \, , \\
&&\int^1_0 dx \int d^2 p'_\perp \frac{h'_0h''_0}{(1-x)\N'_1 \N''_1}
\frac{1}{D'} C^{(1)}_1 \nonumber\\ 
&&= \int^1_0 dx \int d^2 p'_\perp \frac{h'_0h''_0}{(1-x)\N'_1 \N''_1}
\frac{1}{D''} C^{(1)}_1 =0 \, , \\
&&\int^1_0 dx \int d^2 p'_\perp \frac{h'_0h''_0}{(1-x)\N'_1 \N''_1}
\frac{1}{D'} N'_1 C^{(1)}_1 \nonumber\\
&&= \int^1_0 dx \int d^2 p'_\perp \frac{h'_0h''_0}{(1-x)\N'_1 \N''_1}
\frac{1}{D''} N''_1 C^{(1)}_1 =0 \, , 
\eea
\bea
&&\int^1_0 dx \int d^2 p'_\perp \frac{h'_0h''_0}{(1-x)\N'_1 \N''_1}
\frac{1}{D'} \Big[ (q \cdot P +q^2)C^{(2)}_1  \nonumber\\
&&+(P^2 +q \cdot P)B^{(2)}_1 \Big]  \nonumber\\  
&&=\int^1_0 dx \int d^2 p'_\perp \frac{h'_0h''_0}{(1-x)\N'_1 \N''_1}
\frac{1}{D''}\Big[ (q \cdot P -q^2)C^{(2)}_1 +(P^2 -q \cdot P)B^{(2)}_1 \Big]    =0 \, .
\nonumber\\
\eea

We shall not consider the relations that can be derived from the consistency condition
$H^D_1 =0$,
but we shall assume that
\bea
\int^1_0 dx \int d^2 p'_\perp \frac{h'_0h''_0}{(1-x)\N'_1 \N''_1} \; \frac{1}{D' D''} \; Y &=& 0 \, , \\
\int^1_0 dx \int d^2 p'_\perp \frac{h'_0h''_0}{(1-x)\N'_1 \N''_1} \; \frac{1}{D' D''} \; C^{(1)}_1 &=& 0 \, .
\eea
These relations cannot be derived directly but they are a natural generalization of Eqs.(2.35)-(2.38).

The consistency relations (2.35)-(2.42) can be interpreted as an account of the effect of
zero modes in terms of light-front integrals. These results are very useful for practical
calculations. For example, the functions $C^{(1)}_1$, $Y$ and $C^{(2)}_1$ contained in the
form factor $F_2(q^2)$ can be determined by means of Eqs.(2.38) and (2.40)-(2.42).
Consequently, the form factors $F_i(q^2)$ can now be predicted uniquely.

\section{Calculation of form factors that require zero mode contributions}
\setcounter{equation}{0}
\hspace{0.5cm} There is one class of form factors like the electromagnetic
form factor of a pseudoscalar meson, $V$ and $A_2$ for transitions between vector
and pseudoscalar mesons, and the pseudoscalar coupling constant $f_P$
that are free of $C$-functions. These quantities are predicted unambiguously
in 1-loop order by the standard light-front approach.
Another class of form factors like $A_1$ and the vector coupling constant $f_V$
depend on $C$-functions. For their reliable prediction in 1-loop order the
associated zero mode contribution must be known. 
Fortunately, there are only two combinations of $C$-functions, namely $C^{(1)}_1$
and $Y$, that are common to all $C$-dependent form factors considered in this
work. Therefore, the consistency relations (2.35)-(2.42) for the functions
$C^{(1)}_1$ and $Y$, which we have derived for transitions between $q\qb$ vector
mesons, can be used also in other processes to determine
uniquely those form factors that are affected by zero modes. We shall
calculate $A_1(q^2)$ and $f_V$ in this manner; the
corresponding formulas in the SLF and CDKM schemes will be derived in
Appendix C.  

\subsection{The axial form factor $A_1(q^2)$ for transitions between \\
pseudoscalar 
and vector mesons}
\hspace{0.5cm} We shall consider the matrix element of the axial vector current
$A^\m=\bar{q}'' \g^\m \g_5 q'$ between an initial pseudoscalar and a final vector meson state of 4-momentum,
mass and helicity $(P',M')$ and $(P'',M'',h'')$, respectively. Using the same notation
as in Eq.(2.1) it is represented as
\be
G^\m_{h''} = \bra P'',h''|A^\m|P' \ket =
\essm\; f(q^2)+P^\m \; \ess \cdot P \; a_+(q^2)+q^\m \; \ess \cdot P \; a_-(q^2) \, ,
\ee 
where the form factors defined in Eq.(3.1) are related to the convention
used most frequently by
\bea
A_1 (q^2) &=& -(M'+M'')^{-1} f(q^2 ) \, , \nonumber \\
A_2 (q^2 ) &=& (M'+M'')a_+ (q^2 ) \, . 
\eea

This matrix element is given in the 1-loop approximation, corresponding to the
diagram of Fig.1a, as a light-front integral, which we shall denote by $\hat{G}^\m_{h''}$
\be
\hat G^\m_{h''} = \frac{N_c}{16\pi^3}\int^1_0 dx \int d^2 p'_\perp
\frac{\essa tr \left\{ \G' (-\not p_2+m_2) \G''_\a
(\not p''_1 +m''_1) \g^\m \g_5 (\not p'_1 +m'_1) \right\}
}{(1-x) \N'_1 \N''_1} \, .
\ee
The pseudoscalar vertex operator for a $^1S_0$-state meson is \cite{jaus1}
\be
\G' = h'_0 \g_5 \, ,
\ee
and with the vector vertex operator (2.13) the amplitude $\hat{G}^\m_{h''}$ can
be rewritten as
\be
\hat{G}^\m_{h''} = \frac{N_c}{16\pi^3} \int^1_0 dx \int d^2 p'_\perp \frac{h'_0h''_0}
{(1-x) \N'_1 \N''_1} 
  \Big( \hat{A}^\m +\frac{1}{D''}\hat{B}^\m  \Big) \, ,
\ee
where
\bea
\hat{A}_\m&=& -\essa tr\left[\g_\a (\not p''_1+m''_1)\g_\m \g_5 (\not p'_1+m'_1)\g_5 (-\not p_2+m_2)
\right] \, , \nonumber\\
\hat{B}_\m&=& \ess \cdot (p''_1 -p_2) \; tr\left[(\not p''_1+m''_1)\g_\m \g_5 (\not p'_1+m'_1) 
\g_5 (-\not p_2+m_2) \right] \, . \nonumber\\
\eea

Based upon the tensor decomposition of Appendix A the traces
have been represented in Appendix C;
the result for the light-front integral is now given by
\be
\hat{G}^\m_{h''} =\essm \; f(q^2)+P^\m \; \ess \cdot P \; a_+(q^2) +q^\m (\cdots)+\o^\m(\cdots) \, .
\ee
We have again used the identity (2.20) to absorb the contribution due to the term
$\frac{1}{\o \cdot P} P^\m \; \ess \cdot \o$ into the form factor $f(q^2)$ and have
omitted the $\ve$-tensor term in Eq.(3.7).
The form factor $a_+(q^2)$ is free of zero mode contributions and we copy from Ref. \cite{jaus3}:
\bea
a_+ (q^2) &=& \frac{N_c}{16\pi^3} \int ^1_0 dx \int d^2 p'_\perp
\frac{2h'_0 h''_0}{(1-x)\N'_1 \N''_1}
\bigg\{ (2x-1) \left[(1-x)m'_1 +xm_2 \right]  \nonumber \\
& & -\left[2xm_2 +m''_1 +(1-2x)m'_1 \right] \frac{p'_\perp q_\perp}
{q^2} \nonumber \\
& &  -2 \frac{(1-x)q^2 +p'_\perp q_\perp}{(1-x)q^2 D''}
\Big(p'_\perp p''_\perp + \left[xm_2 +(1-x)m'_1 \right]
\left[xm_2 -(1-x)m''_1 \right] \Big) \bigg\} \, , \nonumber\\
\eea
while $f(q^2)$ is
\bea
f(q^2) &=& \frac{N_c}{16\pi^3} \int ^1_0 dx \int d^2 p'_\perp
\frac{h'_0 h''_0}{(1-x)\N'_1 \N''_1} \Bigg\{ 2x(m_2 -m'_1)(M'^2_0 +M''^2_0 )
-4xm''_1 M'^2_0  \nonumber \\
& &  +2(1-x)m'_1 q\cdot P +2m_2 q^2 -2xm_2 (M'^2 +M''^2 ) \nonumber\\
& & +2(m'_1 -m_2 )(m'_1 +m''_1 )^2  \nonumber \\
& &  -8(m'_1 -m_2 ) A^{(2)}_1 +2(m'_1 + m''_1 )(q^2 +q\cdot P) \frac{p'_\perp q_\perp}{q^2}
     -8(m'_1 -m_2 ) B^{(2)}_1 \nonumber \\
& & +\frac{4}{D''}
\Bigg[ A^{(2)}_1 \Big(2xM'^2 +2xM'^2_0-q^2 -q\cdot P 
 -2(q^2 +q\cdot P) \frac{p'_\perp q_\perp }{q^2} \nonumber\\
&& -2(m'_1 -m''_1 )(m'_1 -m_2 ) \Big) \nonumber\\
&& -2B^{(3)}_3 + \Big(M'^2 +M''^2 -q^2 +2(m'_1-m_2)(m''_1 +m_2) \Big) B^{(2)}_1 \nonumber\\ 
&& -m'_1 m''_1 C^{(1)}_1 -Y \Bigg] \Bigg\} \, ,
\eea
where $A^{(2)}_1$, $B^{(2)}_1$ and $C^{(1)}_1$ are defined in Appendix A.
Note that exactly the same result for $f(q^2)$ can be derived from the plus
component of Eq.(3.5), for the longitutinal decay mode, $\hat{G}^+_0$, without
using the decomposition (3.7).

The functions $C^{(1)}_1$ and $Y$, Eq.(2.26), are associated with zero mode effects, and it
is remarkable that there appear the same groups of $C$-functions in Eq.(3.9) as in the
consistency relations of the previous section. In particular, Eqs.(2.37) and (2.38) can
be used to determine the values of $C^{(1)}_1$ and $Y$ in Eq.(3.9):
\be
C^{(1)}_1 \doteq 0 \qquad\mbox{and}\qquad Y \doteq 0 \, .
\ee
The unique result for the form factor $f(q^2)$ is given by Eqs.(3.9)-(3.10).

In our former work \cite{jaus1} we have determined $f(q^2)$ in the framework of an
explicitly covariant model calculation, which we have emphasized not to be realistic.
The difference between that result and Eqs.(3.9)-(3.10) for $f(q^2)$ is given 
by the contribution due to the $B$-functions in Eq.(3.9); the latter vanishes exactly
if the asymmetric vertex functions of the covariant model of Ref. \cite{jaus1} are used.
Therefore, both results are consistent.

In the standard light-front formalism only the amplitude $\hat{G}^+_{h''}$ is used
with the additional requirement that all quarks are on their respective mass shells.
The corresponding expression for $f(q^2)$ has been derived long ago in Ref. \cite{jaus3} 
by means of a different technique and is given again in Appendix C, together with the
expression that is obtained in the CDKM approach of Ref. \cite{karmanov1}.

\subsection{The vector decay constant $f_V$}
\hspace{0.5cm}The vector decay constant $f_V$ is defined by the matrix
element of the vector current
\be
g^\m_h =\bra 0 | V^\m | P ,h \ket =\en^\m  \sqrt{2} f_V \, .
\ee
The matrix element can be represented in 1-loop order by a light-front
momentum integral, which we shall denote as $\hat{g}^\m_h$ :
\be
\hat{g}^\m_h = \frac{N_c}{16\pi^3} \int ^1_0 dx \int d^2 p'_\perp
\frac{h'_0}{(1-x)\N'_1 } s^\m \, ,
\ee
where
\be
s_\m =\en^\n tr \left\{ \g_\m (\not p'_1 +m'_1)
\left[\g_\n - \frac{1}{D'}(p'_1 -p_2)_\n  \right]
(-\not p_2 +m_2 ) \right\} \, , 
\ee
and we have used the vertex operator for $^3S_1$-state mesons, Eq.(2.13).
An explicit representation of the trace can be found in Appendix C.

The decomposition of the integral $\hat{g}^\m_h$, Eq.(3.12), 
into 4-vectors depends on the light-front in the following
way
\be
\hat{g}^\m_h = \sqrt{2} \left\{ \en^\m f_V 
+ \frac{1}{(\o \cdot P)^2} \en \cdot \o \; \o^\m \; r_V \right\} \, ,
\ee
where the term proportional to $r_V$ is spurious. 
We have used an identity similar to Eq.(2.20)
\bea
\frac{1}{\o \cdot P} P^\m \; \en \cdot \o &=& \en^\m + \o^\m \; P^2 \; \frac{\en \cdot \o }{(\o \cdot P)^2} 
\nonumber\\
& &+\frac{1}{(\o \cdot P)^2} \; \ve^{\m \a \b \n}P_\a \o_\b  \; \ve_{\n \t \s \g}\en^\t  
P^\s \o^\g \, ,
\eea
in order to remove the term proportional to $P_\m $, which violates current conservation,
and absorb its contribution into the decay constant $f_V$. Again we neglect the $\ve$-tensor term. 
The result for $f_V$ is
\bea
f_V = \frac{N_c}{8\pi^3} \int ^1_0 dx \int d^2 p'_\perp
\frac{\sqrt{2} h'_0}{(1-x) \N'_1} \Bigg\{ xM'^2_0 -m'_1 (m'_1 -m_2 )
-p'^2_\perp + 2B^{(2)}_1 \nonumber\\
+ \frac{1}{D'}\Big[(m'_1+m_2)( p'^2_\perp -2B^{(2)}_1)+m'_1 C^{(1)}_1 \Big] \Bigg\} \, .
\eea
The standard procedure uses the plus 
component of Eq.(3.12) for the longitutinal decay mode to evaluate 
the vector decay constant and reproduces the result (3.16). In Eq.(3.16) we
have used the same symbols for $B^{(2)}_1$ and $C^{(1)}_1$ as in Appendix A,
though the limit $q^2=q \cdot P=0$ must be taken.

The equation for $f_V$ contains a $C$-function and we use the consistency relation (2.39)
in order to determine the value of $C^{(1)}_1$ in Eq.(3.16):
\be
C^{(1)}_1 \doteq 0 \, .
\ee
The unique result for $f_V$ is given by Eqs.(3.16)-(3.17).

Eqs.(3.16)-(3.17) and the result for $f_V$ found in Ref. \cite{jaus1} differ
again by the contribution due to the $B$-functions in Eq.(3.16), which vanishes
in the covariant model.

The standard light-front expression for the vector decay constant can
be derived from the amplitude $\hat{g}^+_0$ for on-shell quarks \cite{jaus4}, and is
given in Appendix C together with the formula obtained in the CDKM scheme of Ref. \cite{karmanov1}.

\section{Applications}
\setcounter{equation}{0}
\hspace{0.5cm}In Refs. \cite{jaus2,jaus3,jaus4} we have used the standard
light-front formalism to investigate electroweak properties of light and
heavy mesons, and found good agreement with the experimental data. Especially
the predictions obtained in Refs. \cite{jaus4,choi1,choi2} for light
pseudoscalar and vector mesons, for which a large amount of precise
information is available, are remarkably good. The approach of 
Refs. \cite{jaus4,choi1} is based on a Gaussian model wave function,
which is parametrized in terms of adjustable constants, while \cite{choi2}
uses a Gaussian wave function as a trial function of the variational principle
for a QCD-motivated Hamiltonian. The main difference in the best fit 
parameters used in \cite{jaus4,choi1,choi2} was the constituent quark
masses, i.e., $m_u=m_d=m=250$ MeV, $m_s=370$ MeV in \cite{jaus4,choi1},
and $m=250$ MeV, $m_s=480$ MeV in \cite{choi2} for a confining
harmonic oscillator potential.

We expect that the refinements proposed in the present work will lead to
only minor modifications of this general picture. Therefore, we are
mainly interested in a comparison of certain quantitative predictions of
our approach with those of the covariant CDKM scheme and the standard
light-front formalism. For this purpose we shall use the constituent masses
$m=250$ MeV, $m_s=370$ MeV, however, it is interesting to calculate also
the predictions of our formalism for the large s-quark mass $m_s=480$ MeV.
Once the values of the constituent masses are given, the corresponding values
of the wave function parameters $\b$ can be fixed by using the central values of 
the measured decay
constants \cite{PDG} $f_\pi = 92.4 \pm 0.2$ MeV, 
$f_\r /M_\r = 150 \pm 3.6$ MeV, $f_K =113.4 \pm 1.1$ MeV, and  
$f_{K^\ast}/M_{K^\ast} = 153 \pm 3$ MeV. The value for $f_\r /M_\r$ is an average
of the values for $\r^{\pm}$ ($f_{\r^{\pm}} /M_{\r^{\pm}}=147.1 \pm 0.88$ MeV from $Br(\t \to \r \n)
= 0.250 \pm 0.003$) and $\r^0$ ($f_{\r^0} /M_{\r^0}=152.7 \pm 3.5$ MeV from $e^+ e^-$ annihilation).
The value for $f_{K^\ast}/M_{K^\ast}$ can be derived from the branching ratio
$Br(\t \to K^\ast \n) = 0.0129 \pm 0.0005$.
An experimental value for the decay constant $f_D$ of the charm meson 
has been found but with very large errors \cite{PDG}.   
For the purpose of our calculation we
shall use the values 
$\sqrt{2} f_D = 203$ MeV, and
$\sqrt{2} f_{D^\ast}/M_{D^\ast} =238$ MeV. The value for $\sqrt{2} f_D$ [in MeV] is
consistent with the average lattice data obtained in quenched [203(14)] and
unquenched [226(15)] approximations from the recent review \cite{lat1}, and the
QCD sum rule predictions [203(23)], [195(20)] from Refs. \cite{QCD11,QCD22}, respectively.
The value for $\sqrt{2} f_{D^\ast}/M_{D^\ast}$ has been determined with the central
value of the ratio $ f_{D^\ast}/(f_D M_{D^\ast}) = 1.17 \pm 0.06$, which can be 
extracted from the results of lattice calculations presented in \cite{lat2,lat3}.  
We have listed our values
for masses and wave function parameters in Table I.

\begin{table}[h]
\begin{center}
\begin{tabular}{|l|l|l|l|l|l|}
\hline
$(q \bar{Q} )$ meson & $m_Q$ [GeV] & $\b_{P}$ [GeV] & $\b_{V}$ [GeV] & $\b^{SLF}_V$ [GeV]&
$\b^{CDKM}_V$ [GeV] \\  \hline
 $\pi ,\rho$   &  0.25  &  0.3194  & 0.280 & 0.316 & 0.262  \\ \hline
 $K, K^\ast$   &  0.37  &  0.3951  & 0.296 & 0.327 & 0.280    \\ 
               &  0.48  &  0.3629  & 0.294 &        &          \\ \hline
 $D,D^\ast$    &  1.50  &  0.4987  & 0.455  & 0.478 & 0.434    \\ \hline
 
\end{tabular}
\caption{Quark masses $m_Q$ and wave function parameters $\b_{P}$ and
$\b_{V}$ for $(q,\bar{Q} )$ pseudoscalar and vector mesons, which we use
in this work. The wave function parameters to be used in the SLF and CDKM
schemes have been denoted by $\b^{SLF}_V$ and $\b^{CDKM}_V$, respectively.
The light quark mass
is $m_q = m_{u,d} =m = 0.25$ GeV.}
\end{center}
\end{table}

\subsection{Electromagnetic form factors of the $\r$ meson at $q^2=0$.}
\hspace{0.5cm} The transition form factors of a vector meson have been defined
in Eq.(2.1). While the determination of the light-front integrals for the
form factors $F_i(q^2)$ ($i=1,2,3$) is straightforward we shall calculate
them only at $q^2=0$ for $m'_1 =m''_1 =m_2 =m$ and $M'=M''=M$.

For the evaluation of the $p'_\perp$ integrals it is of advantage to change
the momentum variable to $p_\perp =p'_\perp -\frac{1}{2}(1-x)q_\perp$ and
eliminate terms linear in $p_\perp$ by symmetric integration. 

The form factors at $q^2=0$ are 
\bea
F_1(0)&=& \frac{N_c}{16\pi^3} \int^1_0 dx \int d^2 p_\perp \frac{(h_0)^2}
{(1-x) (\N_1)^2} \bigg \{ 2xM^2_0 \nonumber\\
&& +8xA^{(2)}_1 \bigg(1-\frac{4m}{D}+\frac{4m^2-M^2_0}{D^2} \bigg) \bigg \} \, ,
\eea
whith
\[
M^2_0 = \frac{p^2_\perp +m^2}{x(1-x)} \; , \; \N_1 =x(M^2-M^2_0) \, ,
\]
where $h_0$ is given in terms of $M_0$ by Eq.(2.14) and $D=M_0 +2m$.
Note that
\[
1-\frac{4m}{D}+\frac{4m^2-M^2_0}{D^2} =0
\]
for $D=M_0 +2m$.
For the purpose of comparing our method with other approaches we shall use
also a different expression for $D$.

Introducing the longitutinal
momentum $p_3$ and changing variables from $(x,p_\perp)$ to $(p_3,p_\perp)$
with $p^2=p^2_\perp +p^2_3$, $M^2_0=4E^2=4(m^2+p^2)$ and $x=(E+p_3)/M_0$
one can rewrite Eq.(4.1) for $D=M_0 +2m$  as
\be
F_1(0)=\frac{N_c}{(2\pi)^3} \int d^3 p |\phi(M^2_0)|^2 =1 \, .
\ee
This is the normalization condition for the wave function $\phi(M^2_0)$.

The remaining form factors are
\bea
F_2(0)&=&\frac{N_c}{16\pi^3} \int^1_0 dx \int d^2 p_\perp \frac{(h_0)^2}
{(1-x) (\N_1)^2} \bigg \{-4xM^2_0 +2p^2_\perp +2x(1-x)(M^2-M^2_0) \nonumber\\
&& -8\Big[(1-x)B^{(2)}_1+A^{(2)}_1 \Big]
\bigg(1-\frac{4m}{D}+\frac{4m^2-M^2_0}{D^2} \bigg) \nonumber\\
&&-\frac{4m}{D}\Big[ x(1-x)(M^2-M^2_0)
+p^2_\perp \Big]
-\bigg(\frac{d}{dM_0}\frac{1}{D}\bigg)\frac{2m}{M_0} (M^2-M^2_0) p^2_\perp \nonumber\\ 
&&+\frac{4}{D^2} \Big[ 2xM^2 p^2_\perp
+x (M^2-M^2_0) \Big( (1-x)M^2-xM^2_0 \Big) \nonumber\\
&& -x \Big( (1-x)M^2-xM^2_0 \Big)^2 \Big] \bigg \} \, , \\ 
F_3(0)&=& 2M^2 \, \frac{N_c}{16\pi^3} \int^1_0 dx \int d^2 p_\perp \frac{(h_0)^2}
{(1-x) (\N_1)^2} \bigg \{ \frac{4(1-x)m}{D}-\frac{2mp^2_\perp}{xM_0 D^2} \nonumber\\ 
& & +\frac{p^4_\perp}{xM^2_0 D^2} \bigg \} \, ,
\eea
where Eq.(4.4) for $F_3(0)$ is valid only for $D=M_0 +2m$.
The form factors $F_1(0)$ and $F_3(0)$ are free of $C$-functions and are predicted
unambiguously.
Only $F_2(0)$ depends on $C^{(1)}_1$ and $Y$, and we have used the consistency relations
$C^{(1)}_1 \doteq 0$ and $Y \doteq 0$, Eqs.(2.38), (2.41), (2.42).

The specific formulas
in the SLF and CDKM schemes for $D=M_0 +2m$  differ from Eq.(4.3). In the standard 
light-front formalism
$F_2(0)$ can be extracted uniquely from the amplitudes $\hat{G}^+_{h'' h'}$, since the angular 
condition at zero-momentum transfer is fulfilled
for the vector vertex operator (2.13) (e.g. Ref.\cite{cardarelli}). This result can be derived
also from Eqs.(2.24)-(2.29) for on-shell quarks and we find
\be
\D^{SLF}(0)=\frac{1}{M} H^{SLF}_1(0)=0 \, .
\ee
For the determination of $F^{SLF}_2(0)$ we can apply the same method that we have used
for the evaluation of $f^{SLF}(0)$. The result is 
\be
F^{SLF}_2(0)=\frac{N_c}{16\pi^3} \int^1_0 dx \int d^2 p_\perp \frac{(h_0)^2}
{(1-x) (\N_1)^2} \bigg \{-4xM^2_0-\frac{4m}{D}(1-2x)M_0 M \bigg \} \, .
\ee 
In spite of Eq.(4.5), $F^{SLF}_2(0)$ contains spurious contributions related to
unresolved zero mode effects.

If one uses the helicity amplitude $\hat{G}^\perp_{++}$ for on-shell quarks, i.e.,
again avoiding the longitutinal modes, or the representation of the traces (2.18)
given in Appendix B, omitting all $\o$-dependent terms,
one finds the same expression as in the
CDKM scheme of Refs.\cite{karmanov1,melikov1}:
\bea
F^{CDKM}_2(0)&=&\frac{N_c}{16\pi^3} \int^1_0 dx \int d^2 p_\perp \frac{(h_0)^2}
{(1-x)x (\N_1)^2} \bigg \{2p^2_\perp-2xM^2_0-\frac{4m}{D}p^2_\perp \bigg \} \nonumber \\
&=&\frac{N_c}{16\pi^3} \int^1_0 dx \int d^2 p_\perp \frac{(h_0)^2}
{(1-x) (\N_1)^2} \bigg \{-4xM^2_0-\frac{4m}{D}(1-2x)M_0^2 \bigg \} \, . \nonumber \\
\eea

The form factors $F_i$ at $q^2=0$ are related to the static multipole moments of
the charged vector meson by
\[
F_1(0)=1 \;  , \; F_2(0)=-\m \; , \; F_1(0)+F_2(0)+F_3(0)=Q \, ,
\]
where $\m$ is the magnetic moment (in units of $e/2M$) and $Q$ the electric 
quadrupole moment (in units of $e/M^2$).
As a reference we shall use the natural magnetic and quadrupole moment of a composite
spin-1 system which have been derived in \cite{BH} in the limit that the system
becomes pointlike:
\be
\m = 2 \; , \; Q=-1 \quad \mbox{(point limit) .} \quad
\ee
Non-zero values for $\m -2$ and $F_3(0)$ define the anomalous moments, dynamical
contributions which are strictly due to internal structure \cite{BH}.

The magnetic and quadrupole moments of the rho meson, denoted as $\m_\r$ and $Q_\r$, 
respectively, can be calculated in the different schemes with Eqs.(4.3)-(4.7).
With $M_{\r}=769 \, MeV$ and the quark model parameters $(m,\b_\r)$ given
in Table I, we find the results collected in Table II.

\begin{table}[h]
\begin{center}
\begin{tabular}{|l|l|l|l|}
\hline
                   &  this work  &  SLF  &  CDKM   \\  \hline
 $\m_\r=-F_2(0)$   &  1.83  &  2.23 & 2.23   \\ \hline
 $F_3(0)$          &  1.16  &  1.04 & 1.23   \\ \hline
 $Q_\r=F_1(0)+F_2(0)+F_3(0)$   &  0.33  & -0.19 &-0.005    \\ \hline

\end{tabular}
\caption{Magnetic and quadrupole moments of the charged rho meson
calculated in this work are compared with the results obtained in the
SLF and CDKM schemes.}
\end{center}
\end{table}

The anomalous quadrupole moment consists of a large contribution due to $F_3(0)$,
which is predicted unambiguously by the standard light-front formalism, and a
small contribution due to the anomalous magnetic moment $\m_\r -2$, which
depends upon the underlying scheme.

The SLF and CDKM values for $\m_\r$ in Table II are consistent with the results 
($\m_\r=2.26$ and $\m_\r=2.35$) obtained in 
Refs. \cite{cardarelli} and \cite{melikov1}, respectively, using a $q \qb$ wave 
function generated by a quark potential and a quark mass $m=0.22 \, GeV$.

The different values for the magnetic moment of the rho are particularly remarkable,
and it is interesting to compare our results with previous estimates for $\m_\r$.
The nonrelativistic SU(6) model, which neglects all interaction effects, predicts
\be
\m_\r = 2 \quad \mbox{(SU(6) value) ,} \quad
\ee
in accordance with (4.8).
This relation is obtained also in the formalism of hard pions with current algebra \cite{singer},
and in the framework of the Lagrangian formulation of the vector meson dominance hypothesis
\cite{kroll} if strong interaction is neglected \cite{samsonov}.
Inclusion of quark confinement leads to deviations from Eq.(4.9), i.e. to an anomalous
magnetic moment. 
In the SLF and CDKM schemes
of the light-front formalism one finds $\m_\r > 2$, in qualitative agreement with the 
results derived in models based upon the QCD Dyson-Schwinger equation \cite{hawes,hecht}.

In contrast to this picture, our light-front approach, which includes the effect of zero modes,
predicts $\m_\r < 2$, which qualitatively agrees with estimates derived from an analysis
of early SLAC and DESY data for charged $\r$ photoproduction \cite{levy}, and the result of
a recent calculation in the framework of QCD sum rules \cite{samsonov} ($\a_s$-corrections
are, however, neglected). We mention also an oldfashioned phenomenological approach
to estimate $\m_\r$ : The nonrelativistic SU(6) model not only predicts the magnetic
moment of the rho, we write the result as $\m_\r = \m$, but also the rate of the radiative
decay of vector mesons, for example $\o \to \pi + \g$. This rate can be expressed in
terms of $\m$ as (e.g. Ref. \cite{kokkedee})
\[
\G (\o \to \pi \g) = \frac{1}{3 \pi} \bigg( \m \frac{e}{2M_\r} \bigg)^2 
\bigg ( \frac{M^2_\o -M^2_\pi}{2M_\o} \bigg )^3.
\]
We have omitted the overlap integral of the product of pion and $\o$($\r$) wave functions,
introduced by Isgur in Ref. \cite{isgur}, since it is very close to 1 
(if the parameters of Table I are used).
With $\G (\o \to \pi \g) = 0.72 \pm 0.04 \, MeV$, Ref.\cite{PDG}, and treating $\m$ as a free parameter,
one finds $\m_\r = 1.79 \pm 0.05$, close to our value given in Table II.

In order to gain some insight into the physical mechanism that determines the size of the
magnetic moment of the rho, we shall use the toy model, which we analyzed in Ref. \cite{jaus1},
as a guide. Based upon a special class of meson vertex functions
we have shown in \cite{jaus1} that there exists an exact
correspondence between the explicitly covariant 4-dimensional and the
light-front calculations in 1-loop order. While the
vertex functions used in the approach \cite{jaus1} are not symmetric
in the 4-momenta of the constituent quarks, and can hardly be
considered a realistic approximation of the  $q\qb$ bound state, we
expect that the general features of a covariant formalism are reproduced
correctly.

If we use the vector vertex operator (2.13), the toy model  helicity amplitude is
given by Eq.(2.17) with the asymmetric vertex function
\[
h'_0 = \frac{Z'}{\N'_1 - \L^2+m^2} \quad \mbox{(monopole form of the toy model)},
\]
where $\L$ is a cut-off parameter. A similar equation holds for $h''_0$, and we
choose $D'=D''=D =M_\r +2m$. The normalization condition $F_1(0)=1$, where $F_1(0)$
is given by Eq.(4.1), fixes the constant $Z'$, and the toy model value, denoted by 
$\m_\r^{toy}$,
can be computed from Eq.(4.3) for $F_2(0)$. Note that the consistency conditions
for C-functions that have been used for the derivation of the formula (4.3) are
automatically fulfilled in the toy model.
For the numerical calculation we use
the same parameters as in Ref. \cite{ji3}, $m=0.43 \, GeV$ and $\L=1.8 \, GeV$, and include
first only the $\g$-term of the vector vertex operator, and second the full vertex
structure of Eq.(2.13). We give the results for $\m_\r^{toy}$ together with those
obtained in this work:

\[
(\m_\r^{toy},\m_\r^{this work}) = 
\left\{ \begin{array}{ll}
(2.12,2.17)
&({\rm with \; only \; the \; first \; term \; of \; the \; vertex \; (2.13)})\\
(1.79,1.83)
&({\rm with \; both \; terms \; of \; the \; vertex \; (2.13)}).\\
\ea
\right. 
\]
Obviously, the predictions of both approaches are in qualitative agreement.
This calculation indicates that the detailed structure of the vector vertex
influences the size of $\m_\r$, and that the second term of the structure (2.13)
is responsible for the result $\m_\r < 2$. We emphasize that the toy model is
explicitly covariant and automatically includes zero mode contributions, while
the method which we have presented in this work accounts for the effect of
zero modes in terms of specific consistency relations. The SLF and CDKM schemes also
use the full structure of the vector vertex (2.13), but ignore zero mode effects,
which might be the reason for the result $\m_\r > 2$.

\subsection{The transition $V \to P + \pi^0$.}
\hspace{0.5cm} 
In Ref. \cite{jaus2} we have investigated  the pionic decays $\r^+ \to \pi^+ \pi^0$,
$K^{\ast +} \to (K\pi )^+$ and $D^{\ast +} \to (D\pi)^+$,
and have calculated the coupling constant
$g_{VP\pi^0}$ by means of a soft pion theorem due to Das, Mathur and Okubo \cite{das}
\be
4 | g_{VP\pi^0} | f_\pi = | f(0)-(M^2_V -M^2_P)a_+ (0)| 
\ee
where $a_+(0)$ and $f(0)$ are given by Eqs.(3.8) and (3.9), and the SLF and CDKM
expressions for $f(0)$ by Eqs.(C.3)-(C.4).
The pion decay constant
is $f_\pi = 92.4 \, MeV$.
(Note, that if this relation refers to $g_{\r^+ \pi^+ \pi^0}$ the factor 4
in eq.(4.10) must be replaced by the factor 2.) 
The rate for the decay $V \to P + \pi$ is given by
\be
\G = g_{VP \pi}^2 \, p^3 /(6 \pi M_V^2 ),
\ee
\[
p^2 = \left[ M_V^2 - (M_P + M_\pi )^2 \right] \left[ M_V^2 - (M_P - M_\pi)^2 \right]/(4M_V^2 ) \quad .
\]
It might be interesting to compare the coupling constants predicted by our method, and by the SLF and
CDKM schemes, with the data, and we have collected the corresponding numerical results in 
Table III. The coupling constants for different charge states are related by
\bea
g_{\r \pi \pi} &\equiv& g_{\r^+ \pi^+ \pi^0},  \nonumber \\
g_{K^\ast K \pi} &\equiv& \sqrt{3} g_{K^{\ast +}K^+ \pi^0}=\sqrt{3}g_{K^{\ast +}K^0 \pi^+}/\sqrt{2},
 \nonumber \\
g_{D^\ast D \pi} &\equiv& g_{D^{\ast +} D^+ \pi^0} = g_{D^{\ast +} D^0 \pi^+}/ \sqrt{2}.
\nonumber
\eea
The $D^\ast D \pi$ coupling constant is usually expressed in terms of a universal strong 
coupling constant $g$, for the decay of a heavy vector meson into a heavy pseudoscalar
meson and a pion, with
\[
g = 2 g_{D^\ast D \pi} f_\pi /M_{D^\ast}.
\] 
The decay rate $\G_{tot} (D^{\ast +})$ is given by
\[
\G_{tot} (D^{\ast +}) = \G(D^{\ast +} \to (D \pi)^+) + \G(D^{\ast +} \to D^+ \g).
\]
The pionic decay rate $\G(D^{\ast +} \to (D \pi)^+)$ can be calculated with Eq.(4.11).
The (small) radiative decay rate $\G(D^{\ast +} \to D^+ \g)$ depends on the form factor
$V(0)$, which is free of C-functions, and for its evaluation we have used the results
of Ref. \cite{jaus2}.

\begin{table}[h]
\begin{center}
\begin{tabular}{|l|l|l|l|l|}
\hline
                      & this work & SLF   & CDKM  & experiment  \\  \hline
 $g_{\r \pi \pi}$     &  6.10     & 5.62  & 10.0 & $6.06 \pm 0.01$ \cite{PDG} \\ \hline
 $g_{K^\ast K \pi}$   & 5.43      & 5.20  & 9.41  & $5.57 \pm 0.03$ \cite{PDG}  \\ 
                      & 5.85      &       &       &                             \\ \hline
 $g_{D^\ast D \pi}$   & 6.24      & 6.22  & 11.16 & $6.3 \pm 0.1 \pm 0.7$ \cite{cleo}   
\\ \hline
 $g$                  & 0.57      & 0.57  & 1.03  & $0.59 \pm 0.01 \pm 0.07$ \cite{cleo}
\\ \hline
 $\G_{tot} (D^{\ast +})$ [keV] & 93.0 & 92.4 & 295.7 & $96 \pm 4 \pm 22$ \cite{cleo}
  \\ \hline
 
\end{tabular}
\caption{Coupling constants for $V \to P \pi$ decays, and the total rate for the 
decay of the charged $D^\ast$ meson. The parameters of Table I have been used.}
\end{center}
\end{table}

A discussion of $g$ in the SLF scheme can be found in Ref. \cite{xu}. For no Melosh rotation
and $\b_D =\b_{D^\ast}$ one would get $g=1$ as in the nonrelativistic quark model \cite{NRQM}.
In the framework of the light-front formalism the Melosh rotation generates the relativistic
spinor structure of the vertex operators (2.13) and (3.4) and generally one obtains $g < 1$.
The precise value of $g$ depends upon the relative size of the constituent masses. For example,
for a small (yet unrealistic) constituent mass $m_{u/d} = 100 \, MeV$ one finds 
the unlikely value $g \simeq 0.3$ (this work and SLF). If the constituent mass approaches
zero $g$ gets very small.
For the set of parameters given in Table I
the agreement of the predictions of our approach and of the SLF scheme with the data
is satisfactory and within the range expected from the effect of an underlying approximate
chiral symmetry. The predictions based upon the CDKM scheme are surprisingly large, which
seems to indicate that the prescriptions used for the calculation of $f^{CDKM}(q^2)$ are
probably too restrictive, i.e. we suspect that either the on-mass-shell prescription or the
prescription that all $\o$-dependent contributions should be omitted, is responsible
for these results. In this context it may be interesting to note that within the
CDKM scheme it has been found for the deuteron electrodisintegration \cite{math1} that
instantaneous contributions can be large, and for nucleon form factors \cite{math2} that
$\o$-dependent structures can give contributions to physical form factors.

Of particular importance is the recent measurement of the total decay rate of the charged 
$D^\ast$ meson by the CLEO collaboration \cite{cleo}, since the coupling constant
$g_{D^\ast D\pi}$ is of considerable theoretical interest and
has been studied extensively in past work (an overview and a list
of references can be found in  \cite{yaouanc,singer2}). The range of published
values for $g_{D^\ast D\pi}$ is quite large;
we only mention the recent
lattice QCD result $g_{D^\ast D\pi}=6.6 \pm 0.8^{+0.4}_{-0.7}$ \cite{abada}, which is 
consistent with the data (and with our value given in Table III), and the results based 
upon various QCD sum rule techniques \cite{QCD1}
(for example Ref. \cite{QCD2} predicts $g_{D^\ast D\pi}=3.7 \pm 1.1$, i.e.,
$g = 0.34$ close to the ``unlikely'' light-front value), which are much
smaller than the data. However, in a recent reanalysis of the $D^\ast D\pi$ coupling in the
framework of the light-cone QCD sum rule \cite{kim} it was pointed out how agreement
with the data could be attained (another attempt has been published in \cite{lat4}). 

It would be interesting to explore the connection
between the light-front formalism and the  light-cone QCD sum rule approach.

\subsection{Determination of current quark masses.}
\hspace{0.5cm} 
The masses of the quarks in the light-front formalism are the constituent masses.
Even in this framework it is possible to obtain an estimate of the 
current quark masses from the Ward identity for the axial current. It can be
derived starting from the matrix element of the axial current between the vacuum
and a $q \qb$ meson
\be
\bra 0 |\bar{q}'' \g_\m \g_5 q'  | P  \ket = i P_\m \sqrt{2} f_P,
\ee
and is given by \cite{leutwyler1}
\be
-(\hat{m}_{q'}+\hat{m}_{q''})\bra 0 |\bar{q}''  \g_5 q'  | P  \ket =
i \sqrt{2} f_P M^2_P.
\ee
This relation involves the current quark masses $\hat{m}_q$ and is exact, except 
for electroweak corrections.

In analogy to Eq.(4.12) we define the coupling constant $g_P$ of the pseudoscalar
density as
\be
\bra 0 |\bar{q}''  \g_5 q'  | P  \ket = -i \sqrt{2} g_P,
\ee
and Eq.(4.13) can be rewritten as an equation for the quark masses
\be
\hat{m}_{q'}+\hat{m}_{q''}=M^2_P f_P /g_P.
\ee
It is not possible to measure the coupling constant $g_P$, but the formalism presented 
in this work permits a reliable determination of the matrix element (4.14) in terms of
constituent masses $m_q$ and wave function parameters $\b$. Since the free parameters
$(m_q,\b)$ are fixed by comparison with the data, Eq.(4.15) provides an interesting
phenomenological link between the current quark masses $\hat{m}_q$ and experiment.

The equation for the pseudoscalar decay constant $f_P$ can be found e.g. in \cite{jaus1}, 
and the matrix element (4.14)
is given in the 1-loop approximation as a light-front integral; the corresponding
expression for $g_P$ is
\be
g_P=\frac{N_c}{16\pi^3} \int ^1_0 dx \int d^2 p'_\perp
\frac{\sqrt{2} h'_0}{2(1-x)\N'_1 } s,
\ee
where
\bea
s &=& tr \left\{ \g_5 (\not p'_1 +m'_1)\g_5
(-\not p_2 +m_2 ) \right\} \nonumber\\
 &=& 2\Big( M^2_P-N'_1-N_2-(m'_1-m_2)^2 \Big). \nonumber
\eea
With Eqs.(A.3)-(A.4) and(A.13) we find for the trace
\[
s=2 \Big( 2xM'^2_0-2m'_1(m'_1-m_2)+C^{(1)}_1 \Big).
\]
Obviously $g_P$ is associated with zero modes.
The final result for $g_P$ is
\be
g_P=\frac{N_c}{8\pi^3} \int ^1_0 dx \int d^2 p'_\perp
\frac{\sqrt{2} h'_0}{2(1-x)\N'_1 } \Big(xM'^2_0-m'_1(m'_1-m_2)\Big),
\ee
where we have used the consistency relation (2.39) in order to eliminate $C^{(1)}_1$.
Note that the on-mass-shell prescription (all quarks are on their mass shells) gives the same
result for $g_P$; the latter has been derived in this manner in Ref. \cite{kiss}.

If the integral (4.17) is evaluated with the quark model parameters of Table I, we find,
using Eq.(4.15) with the average mass squared $M^2_P=(M^2_{P^0}+M^2_{P^+})/2$:
\[
\hat{m}=\frac{1}{2} (\hat{m}_u+\hat{m}_d)= 4.39 \, MeV \quad \mbox{(for m = 250 MeV)},
\]
\[
\hat{m}_s +\hat{m}=
\left\{ \begin{array}{ll}
91.96 \, MeV
&({\rm for \, m_s=370 \, MeV})\\
106.36 \, MeV
&({\rm for \, m_s=480 \, MeV}).\\
\ea
\right. 
\]
Of particular interest are the estimates for the ratio $\hat{m}_s/\hat{m}$
\[
\hat{m}_s/\hat{m}=
\left\{ \begin{array}{ll}
19.97
&({\rm for \, m_s=370 \, MeV})\\
23.25
&({\rm for \, m_s=480 \, MeV}),\\
\ea
\right. 
\]
which can be compared with the result obtained in chiral perturbation theory at
the next-to-leading order \cite{leutwyler2}
\be
\hat{m}_s/\hat{m}=24.4 \pm 1.5.
\ee 
Evidently only the large value of the constituent mass of the s-quark is consistent
with (4.18).

\section{Concluding remarks}
\hspace{0.5cm} We have proposed in this work a method to account for the effect
of zero modes that are naturally associated with the light-front integrals that
represent the matrix elements of a one-body current. In order to identify that
part of the light-front integral that is influenced by zero modes,
we have decomposed it in terms of the 4-vectors $P$, $q$ and $\o$,
and found that the decomposition coefficients $C^{(m)}_n$ 
must be completed by (unknown) zero mode contributions. 
Fortunately there are only two combinations
of C-functions, namely $C^{(1)}_1$ and $Y$, that are common to all form
factors that are affected by zero modes. With plausible assumptions it
is possible to derive consistency conditions for the functions
$C^{(1)}_1$ and $Y$, which permit an unambiguous determination of the
C-dependent form factors. This method provides a practical scheme to
calculate the effect of zero modes, and constitutes a covariant extension
of the standard light-front formalism.

The correctness of this procedure is tested by the phenomenological
success of the derived form factors. In practical applications it appears 
that the predictions of the SLF approach and the covariant extension proposed
in this work are usually very similar, i.e., the good agreement with the data
found in the SLF framework is confirmed. However, there are problems like the
$q^2$ dependence of the electromagnetic form factors of vector mesons, or the details
of the magnetic moment of the rho that can be treated only
by the refined formalism.

Based upon rules, similar to those established for $q \qb$ mesons, it should be 
possible to determine also zero-mode-affected form factors of $qqq$ baryons. For
example, it is remarkable that the charge form factor of the proton is free of
C-functions, but its magnetic form factor depends on C-functions in terms of
expressions corresponding to $C^{(1)}_1$ and $Y$.

\subsection*{Acknowledgements}
\hspace{0.5cm} I would like to thank Daniel Wyler for helpful discussions.

\begin{appendix}
\section{Appendix: Decomposition of tensors}
\setcounter{equation}{0}
\hspace{0.5cm} 
In order to treat the Lorentz structure of a hadronic
matrix element the authors of Ref. \cite{karmanov1} have developed
a method to identify and separate spurious contributions and to
determine the physical part of the hadronic matrix element. We have
developed an essentially different technique in Ref. \cite{jaus1} to deal
with this problem. It is based on the decomposition of 4-tensors
with regard to $P$, $q$ and $\o$; we shall copy the respective formulas which we
shall need in the present work from Ref. \cite{jaus1}.

The integrand of a light-front integral, like Eq.(2.17), depends on the tensors
$p'_{1 \m}$, $p'_{1 \m}p'_{1 \n}$ and $p'_{1 \m}p'_{1 \n}p'_{1 \a}$. Let us start
with the 4-vector $p'_{1 \m}$ whose decomposition is inferred from
symmetry considerations to be
\bea
\int^1_0 dx \, \int d^2p'_\perp \,  \frac{R'S''}
{(1-x) \N'_1 \N''_1} p'_{1\m} 
=\int^1_0 dx \, \int d^2p'_\perp \, \frac{R'S''}
{(1-x) \N'_1 \N''_1}
\Big( P_\m A^{(1)}_1
+q_\m A^{(1)}_2 \nonumber\\
+\frac{1}{\o \cdot P}\o_\m C^{(1)}_1 \Big) \nonumber \\
\eea
where $R'$ and $S''$ are functions of $M'_0$ and $M''_0$, respectively.

An equation like (A.1) will be written in the following as a relation
between integrands as
\be
p'_{1\m}\doteq P_\m A^{(1)}_1+q_\m A^{(1)}_2+\frac{1}{\o \cdot P}\o_\m C^{(1)}_1.
\ee
The coefficients in (A.1) and (A.2) are given by
\be
A^{(1)}_1=\frac{x}{2} \quad ,\quad
A^{(1)}_2=\frac{x}{2}-\frac{p'_\perp q_\perp}{q^2}  \quad , \quad
C^{(1)}_1=-N_2+Z_2 ,
\ee
where
\be
Z_2=-xM'^2_0+m'^2_1-m^2_2+(1-x)M'^2+(q^2+q\cdot P)\frac{p'_\perp q_\perp}{q^2}.
\ee
Note that only the coefficient which is combined with $\o_\m$, namely
$C^{(1)}_1$, depends
on $p'^-_1$. 

We shall need also the tensor decomposition
\bea
p'_{1\m} p'_{1\n}& \doteq &g_{\m \n}A^{(2)}_1+P_\m P_\n A^{(2)}_2
+(P_\m q_\n+q_\m P_\n)A^{(2)}_3+q_\m q_\n A^{(2)}_4  \nonumber \\
& &+\frac{1}{\o \cdot P}(P_\m \o_\n+\o_\m P_\n)B^{(2)}_1
+\frac{1}{\o\cdot P}(q_\m \o_\n+\o_\m q_\n)C^{(2)}_1 \nonumber \\
& & +\frac{1}{(\o \cdot P)^2}\o_\m \o_\n C^{(2)}_2, 
\eea
where
\bea
A^{(2)}_1&=&-p'^2_\perp-\frac{(p'_\perp q_\perp)^2}{q^2} \quad ,\quad
A^{(2)}_2=(A^{(1)}_1)^2, \nonumber \\
A^{(2)}_3&=&A^{(1)}_1A^{(1)}_2 \quad ,\quad
A^{(2)}_4=(A^{(1)}_2)^2-\frac{1}{q^2}A^{(2)}_1, \nonumber\\
B^{(2)}_1&=&A^{(1)}_1C^{(1)}_1-A^{(2)}_1 \quad ,\quad
C^{(2)}_1=A^{(1)}_2C^{(1)}_1+\frac{q \cdot P}{q^2}A^{(2)}_1, \nonumber \\
C^{(2)}_2&=&(C^{(1)}_1)^2+\left[P^2-\frac{(q \cdot P)^2}{q^2}\right]A^{(2)}_1.
\eea
Since the zero mode term of $xN_2$ vanishes we find the following relation:
\be
xN_2 \doteq x\N_2 =0,
\ee
and there are no zero-mode contributions associated with
$B^{(2)}_1$, which is therefore given by its value at the spectator quark pole:
\be
B^{(2)}_1=\frac{x}{2}Z_2-A^{(2)}_1.
\ee

For the practical applications considered in this work we shall need also
the decomposition of the following tensor product
\bea
p'_{1\m}p'_{1\n}p'_{1\a}&\doteq&(\gmn P_\a +\gma P_\n +\gna P_\m)A^{(3)}_1
+(\gmn q_\a +\gma q_\n +\gna q_\m)A^{(3)}_2  \nonumber\\
&& +P_\m P_\n P_\a A^{(3)}_3+(P_\m P_\n q_\a+P_\m q_\n P_\a+q_\m P_\n P_\a)A^{(3)}_4 \nonumber\\
&& +(q_\m q_\n P_\a +q_\m P_\n q_\a+P_\m q_\n q_\a)A^{(3)}_5+q_\m q_\n q_\a A^{(3)}_6 \nonumber\\
&&+\frac{1}{\o \cdot P}(P_\m P_\n \o_\a +P_\m \o_\n P_\a +\o_\m P_\n P_\a)B^{(3)}_1 \nonumber\\
&&+\frac{1}{\o \cdot P}\Big[(P_\m q_\n+q_\m P_\n)\o_\a +(P_\m q_\a +q_\m P_\a)\o_\n \nonumber\\
&& +(P_\n q_\a +q_\n P_\a)\o_\m \Big]B^{(3)}_2 \nonumber \\
&&+\frac{1}{\o \cdot P}(\gmn \o_\a+\gma \o_\n+\gna \o_\m)C^{(3)}_1 \nonumber \\
&&+\frac{1}{\o \cdot P}(q_\m  q_\n \o_\a+q_\m \o_\n q_\a+\o_\m q_\n q_\a)C^{(3)}_2\nonumber\\
&& +\frac{1}{(\o \cdot P)^2}(\o_\m \o_\n P_\a +\o_\m P_\n \o_\a +P_\m \o_\n \o_\a)C^{(3)}_3 \nonumber\\
&&+\frac{1}{(\o \cdot P)^2}(\o_\m \o_\n q_\a +\o_\m q_\n \o_\a +q_\m \o_\n \o_\a)C^{(3)}_4 \nonumber\\
&&+\frac{1}{(\o \cdot P)^3}\o_\m \o_\n \o_\a C^{(3)}_5. 
\eea
The coefficients are given by
\bea
A^{(3)}_1&=&A^{(1)}_1A^{(2)}_1 \qquad , \qquad 
A^{(3)}_2=A^{(1)}_2A^{(2)}_1, \nonumber \\
A^{(3)}_3&=&A^{(1)}_1A^{(2)}_2 \qquad , \qquad
A^{(3)}_4=A^{(1)}_2A^{(2)}_2, \nonumber \\
A^{(3)}_5&=&A^{(1)}_1A^{(2)}_4 \qquad , \qquad
A^{(3)}_6=A^{(1)}_2A^{(2)}_4-\frac{2}{q^2}A^{(1)}_2A^{(2)}_1, \nonumber\\
B^{(3)}_1&=&A^{(1)}_1B^{(2)}_1-A^{(1)}_1A^{(2)}_1 \quad , \quad 
B^{(3)}_2=A^{(1)}_1C^{(2)}_1-A^{(1)}_2A^{(2)}_1, \nonumber \\
C^{(3)}_1&=&A^{(2)}_1 C^{(1)}_1 \qquad , \qquad
C^{(3)}_2=A^{(2)}_4 C^{(1)}_1 +2\frac{q\cdot P}{q^2}A^{(1)}_2A^{(2)}_1, \nonumber\\
C^{(3)}_3&=&A^{(1)}_1(C^{(1)}_1)^2-2C^{(3)}_1+\Big(P^2-\frac{(q \cdot P)^2}{q^2} \Big)A^{(1)}_1A^{(2)}_1,
\nonumber\\
C^{(3)}_4&=&A^{(1)}_2(C^{(1)}_1)^2+2\frac{q\cdot P}{q^2}C^{(3)}_1
+\Big(P^2-\frac{(q \cdot P)^2}{q^2} \Big)A^{(1)}_2A^{(2)}_1.
\eea
We do not require the coefficient $C^{(3)}_5$.
Again the coefficients $B^{(3)}_1$
and $B^{(3)}_2$ are not associated with zero-mode contributions and are
given by their values at the spectator quark pole:
\bea
B^{(3)}_1&=&\frac{x}{2}\left(B^{(2)}_1-A^{(2)}_1\right)  \\
B^{(3)}_2&=&\left(\frac{x}{2}-\frac{p'_\perp q_\perp}{q^2}\right) B^{(2)}_1
+\frac{x}{2}\frac{q\cdot P}{q^2}A^{(2)}_1. 
\eea
Useful relations can be derived from the tensor decompositions (A.5) and (A.9):
\be
N'_1 = p'^2_1 -m'^2_1 = x(M'^2-M'^2_0),
\ee
in accordance with Eq.(2.12), and
\bea
N'_1 C^{(1)}_1 &=& 2C^{(3)}_3+6C^{(3)}_1+q^2C^{(3)}_2-m'^2_1C^{(1)}_1+P^2 B^{(3)}_1+2q\cdot P B^{(3)}_2, \\
N''_1 C^{(1)}_1 &=& 2C^{(3)}_3+6C^{(3)}_1+q^2C^{(3)}_2-2q^2C^{(2)}_1 \nonumber\\
&& +(q^2-m''^2_1)C^{(1)}_1+P^2 B^{(3)}_1+2q\cdot P B^{(3)}_2-2q\cdot P B^{(2)}_1 .
\eea
These relations can be combined with Eq.(2.26) for $Y$ to give
\bea
(N'_1+N''_1)C^{(1)}_1&=& 2Y+4C^{(3)}_3+8C^{(3)}_1+(q^2-m'^2_1-m''^2_1)C^{(1)}_1, \\
(N'_1-N''_1)C^{(1)}_1&=& q^2(2C^{(2)}_1-C^{(1)}_1)-(m'^2_1-m''^2)C^{(1)}_1+2q\cdot P B^{(2)}_1.
\eea
In the same manner we find
\bea
N_2C^{(1)}_1 &=& 2C^{(3)}_3+6C^{(3)}_1+q^2C^{(3)}_2-C^{(2)}_2-(q^2+q \cdot P)C^{(2)}_1 \nonumber\\
&& +(M'^2-m^2_2)C^{(1)}_1+P^2 B^{(3)}_1+2q\cdot P B^{(3)}_2-(P^2+q\cdot P)B^{(2)}_1, \nonumber \\
&& \\
N_2 B^{(2)}_1 &=& -C^{(3)}_3-C^{(3)}_1+B^{(3)}_3 ,
\eea
where
\be
B^{(3)}_3 = B^{(2)}_1 Z_2+ \bigg( P^2-\frac{(q\cdot P)^2}{q^2} \bigg) A^{(1)}_1 A^{(2)}_1.
\ee

\section{Appendix: The traces of Eq.(2.18)}
\setcounter{equation}{0}
\hspace{0.5cm} Using the tensor decomposition of Appendix A the traces of Eq.(2.18)
can be represented for $m'_1=m''_1=m_2=m$  as
\bea
A^\m &=& \ess \cdot \es \: P^\m \bigg\{ 8xA^{(2)}_1+(1-x)q^2+x(M'^2_0+M''^2_0) \bigg\} \nonumber\\
&& + (\ess \cdot P)(\es \cdot P)\: P^\m \, 8x \bigg(A^{(2)}_2-A^{(2)}_4+A^{(1)}_2-A^{(1)}_1 \bigg)
\nonumber\\
&& + (\es^\m \: \ess \cdot P+\essm \: \es \cdot P ) \bigg\{16A^{(3)}_1-8A^{(2)}_1-2C^{(1)}_1-4xM'^2_0
\nonumber\\
&& -x(1-x)(M'^2_0+M''^2_0)+x(1-x)(M'^2+M''^2)+2x(M'^2_0-M''^2_0)A^{(1)}_2 \nonumber\\
&& -2q^2A^{(1)}_2-(1-2x)q^2+x(1-2A^{(1)}_2)q\cdot P \bigg\} \nonumber\\
&& +(\es^\m \: \ess \cdot P-\essm \: \es \cdot P ) \bigg\{16A^{(3)}_2-8A^{(2)}_1+x^2(M'^2_0-M''^2_0)
\nonumber\\
&& +x(1-x)q\cdot P+(2A^{(1)}_2-1)(q^2-xM'^2-xM''^2+xM'^2_0+xM''^2_0) \bigg\} \nonumber\\
&& +\frac{1}{\o \cdot P}(\ess \cdot P)(\es \cdot \o)\: P^\m \bigg(16B^{(3)}_1+16B^{(3)}_2-16B^{(2)}_1
   +2C^{(1)}_1 \bigg) \nonumber\\
&& +\frac{1}{\o \cdot P}(\ess \cdot \o)(\es \cdot P)\: P^\m \big(16B^{(3)}_1-16B^{(3)}_2
   +2C^{(1)}_1 \big) \nonumber\\
&& +\frac{1}{(\o \cdot P)^2}(\ess \cdot \o)(\es \cdot \o)\: P^\m \: 16C^{(3)}_3 \nonumber\\
&& +\frac{1}{\o \cdot P}(\es^\m \: \ess \cdot \o +\essm \: \es \cdot \o ) \bigg\{-8C^{(3)}_3
   -4Y+4m^2C^{(1)}_1 \bigg\} \nonumber\\
&& +\frac{1}{\o \cdot P}(\es^\m \: \ess \cdot \o -\essm \: \es \cdot \o ) \bigg\{-2(N'_1-N''_1)C^{(1)}_1
   \bigg\} \nonumber\\
&& +q^\m \{\cdots\} + \o^\m \{ \cdots \} , 
\eea
\bea
B^\m &=& \ess \cdot \es \: P^\m \: 16m xA^{(2)}_1  \nonumber\\
&& + (\ess \cdot P)(\es \cdot P)\: P^\m \: 2m \bigg\{ 8x (A^{(2)}_2-A^{(2)}_4+A^{(1)}_2-A^{(1)}_1 )
   -2(A^{(1)}_2-A^{(1)}_1) \bigg\} \nonumber\\
&& + (\es^\m \: \ess \cdot P+\essm \: \es \cdot P ) \: 2m \bigg\{16A^{(3)}_1-8A^{(2)}_1
\nonumber\\
&& -(A^{(1)}_2-A^{(1)}_1)(q^2-2xM''^2+2xM''^2_0-q\cdot P) \bigg\} \nonumber\\
&& +(\es^\m \: \ess \cdot P-\essm \: \es \cdot P ) \: 2m \bigg\{16A^{(3)}_2-8A^{(2)}_1 \nonumber\\
&& +(A^{(1)}_2-A^{(1)}_1)(q^2-2xM''^2+2xM''^2_0-q\cdot P) \bigg\} \nonumber\\
&& +\frac{1}{\o \cdot P}(\ess \cdot P)(\es \cdot \o)\: P^\m \: 2m \bigg\{16B^{(3)}_1+16B^{(3)}_2-16B^{(2)}_1
   +2C^{(1)}_1 \bigg\} \nonumber\\
&& +\frac{1}{\o \cdot P}(\ess \cdot \o)(\es \cdot P)\: P^\m \: 2m \bigg\{16B^{(3)}_1-16B^{(3)}_2
    \bigg\} \nonumber\\
&& +\frac{1}{(\o \cdot P)^2}(\ess \cdot \o)(\es \cdot \o)\: P^\m \: 32mC^{(3)}_3 \nonumber\\
&& +\frac{1}{\o \cdot P}(\es^\m \: \ess \cdot \o +\essm \: \es \cdot \o ) \: 2m \bigg\{-8C^{(3)}_3
   -4Y+4m^2C^{(1)}_1 \nonumber\\
&& +2N'_1C^{(1)}_1-(q^2+q\cdot P)C^{(1)}_1 \bigg\} \nonumber\\
&& +\frac{1}{\o \cdot P}(\es^\m \: \ess \cdot \o -\essm \: \es \cdot \o )\: 2m \bigg\{2N''_1C^{(1)}_1
   -(q^2-q\cdot P)C^{(1)}_1 \bigg\} \nonumber\\
&& +q^\m \{\cdots\} + \o^\m \{ \cdots \} ,
\eea
\bea
C^\m &=& \ess \cdot \es \: P^\m \: 16m xA^{(2)}_1  \nonumber\\
&& + (\ess \cdot P)(\es \cdot P)\: P^\m \: 2m \bigg\{ 8x (A^{(2)}_2-A^{(2)}_4+A^{(1)}_2-A^{(1)}_1 ) \nonumber \\
&&   +2(A^{(1)}_2+A^{(1)}_1-1) \bigg\} \nonumber\\
&& + (\es^\m \: \ess \cdot P+\essm \: \es \cdot P ) \: 2m \bigg\{16A^{(3)}_1-8A^{(2)}_1
\nonumber\\
&& +(A^{(1)}_2+A^{(1)}_1-1)(q^2-2xM'^2+2xM'^2_0+q\cdot P) \bigg\} \nonumber\\
&& +(\es^\m \: \ess \cdot P-\essm \: \es \cdot P ) \: 2m \bigg\{16A^{(3)}_2-8A^{(2)}_1 \nonumber\\
&& +(A^{(1)}_2+A^{(1)}_1-1)(q^2-2xM'^2+2xM'^2_0+q\cdot P) \bigg\} \nonumber\\
&& +\frac{1}{\o \cdot P}(\ess \cdot P)(\es \cdot \o)\: P^\m \: 2m \bigg\{16B^{(3)}_1+16B^{(3)}_2-16B^{(2)}_1
    \bigg\} \nonumber\\
&& +\frac{1}{\o \cdot P}(\ess \cdot \o)(\es \cdot P)\: P^\m \: 2m \bigg\{16B^{(3)}_1-16B^{(3)}_2
    +2C^{(1)}_1 \bigg\} \nonumber\\
&& +\frac{1}{(\o \cdot P)^2}(\ess \cdot \o)(\es \cdot \o)\: P^\m \: 32mC^{(3)}_3 \nonumber\\
&& +\frac{1}{\o \cdot P}(\es^\m \: \ess \cdot \o +\essm \: \es \cdot \o ) \: 2m \bigg\{-8C^{(3)}_3
   -4Y+4m^2C^{(1)}_1 \nonumber\\
&& +2N''_1C^{(1)}_1-(q^2-q\cdot P)C^{(1)}_1 \bigg\} \nonumber\\
&& +\frac{1}{\o \cdot P}(\es^\m \: \ess \cdot \o -\essm \: \es \cdot \o )\: 2m \bigg\{-2N'_1C^{(1)}_1
   +(q^2+q\cdot P)C^{(1)}_1 \bigg\} \nonumber\\
&& +q^\m \{\cdots\} + \o^\m \{ \cdots \} ,
\eea
\bea
D^\m &=& \ess \cdot \es \: P^\m \: 4A^{(2)}_1 \bigg\{x(8m^2-M'^2_0-M''^2_0)-(1-x)q^2 \bigg\}  \nonumber\\
&& + (\ess \cdot P)(\es \cdot P)\: P^\m \: 4(A^{(2)}_2-A^{(4)}_2+A^{(1)}_2-A^{(1)}_1 ) \bigg\{
   x(8m^2-M'^2_0-M''^2_0) \nonumber \\
&& -(1-x)q^2 \bigg\} \nonumber\\
&& + (\es^\m \: \ess \cdot P+\essm \: \es \cdot P )  \bigg\{4(1-x)A^{(2)}_1(M'^2+M''^2-q^2-8m^2) \nonumber\\
&& -8A^{(2)}_1Z_2+8C^{(3)}_1 \bigg\} \nonumber\\
&& +(\es^\m \: \ess \cdot P-\essm \: \es \cdot P ) \bigg\{4(2A^{(3)}_2-A^{(2)}_1)(q^2-M'^2-M''^2 \nonumber \\
&& +2N_2+8m^2)
 \bigg\} \nonumber\\
&& +\frac{1}{\o \cdot P}(\ess \cdot P)(\es \cdot \o)\: P^\m \bigg\{8B^{(3)}_1(q^2-M'^2-M''^2+8m^2) \nonumber\\
&& +8(B^{(3)}_2-B^{(2)}_1)(q^2-M'^2-M''^2+8m^2+2N_2) \nonumber\\
&& -4(B^{(2)}_1+C^{(2)}_1-C^{(1)}_1)(q^2-N'_1-N''_1)
   \bigg\} \nonumber\\
&& +\frac{1}{\o \cdot P}(\ess \cdot \o)(\es \cdot P)\: P^\m \bigg\{8B^{(3)}_1(q^2-M'^2-M''^2+8m^2) \nonumber\\
&& -8B^{(3)}_2(q^2-M'^2-M''^2+8m^2+2N_2)-4(B^{(2)}_1-C^{(2)}_1)(q^2-N'_1-N''_1)
     \bigg\} \nonumber\\
&& +\frac{1}{(\o \cdot P)^2}(\ess \cdot \o)(\es \cdot \o)\: P^\m \bigg\{8C^{(3)}_3(q^2-M'^2-M''^2+8m^2+2N_2)
\nonumber\\
&& -4C^{(2)}_2(q^2-N'_1-N''_1) \bigg\} \nonumber\\
&& +\frac{1}{\o \cdot P}(\es^\m \: \ess \cdot \o +\essm \: \es \cdot \o ) \: 8C^{(3)}_1
   \bigg\{q^2-M'^2-M''^2+8m^2+2N_2 \bigg\} \nonumber\\
&& +q^\m \{\cdots\} + \o^\m \{ \cdots \} ,
\eea
where we have used the relation (A.7) and $Y$ is given by Eq.(2.26). The trace $D^\m$ contains higher order
momentum contributions which we have not analyzed in detail, but those contributions that are required for
the computation of the form factors $F_1(q^2)$, $F_2(q^2)$ and $F_3(q^2)$ are unambiguously determined by
our approach.

\section{Appendix: Determination of $f(q^2)$ and $f_V$ in the SLF and CDKM schemes }
\setcounter{equation}{0}
\hspace{0.5cm} First we shall derive the formulas for $f(q^2)$.
Using the tensor decomposition of Appendix A the traces of Eq.(3.6) can be represented as
\bea
\hat{A}^\m&=& \essm \Big\{-2m''_1 \Big( M'^2-N'_1-N_2-(m'_1-m_2)^2 \Big)-8(m'_1-m_2)A^{(2)}_1 \nonumber\\
&& -2m'_1 \Big( M''^2-N''_1-N_2-m''^2_1-m^2_2 \Big) \nonumber\\
& & +2m_2 \Big( q^2-N'_1-N''_1-m'^2_1-m''^2_1 \Big) \Big\}
\nonumber\\
&& +P^\m \ess \cdot P \Big\{ -8(m'_1-m_2)(A^{(2)}_2+A^{(2)}_3)+2(m'_1+m''_1)A^{(1)}_2 \nonumber\\
&&+2(5m'_1-m''_1-2m_2)A^{(1)}_1-2m'_1 \Big\} \nonumber\\
&& + P^\m \frac{\ess \cdot \o}{\o \cdot P} \Big\{ 8(m_2-m'_1)B^{(2)}_1 +2(m'_1+m''_1)C^{(1)}_1 \Big\} \nonumber\\
&& +q^\m \{\cdots\} + \o^\m \{ \cdots \} \, , \nonumber\\
\hat{B}^\m&=& \essm \; 4A^{(2)}_1 \Big\{ M'^2+M''^2-q^2-2N_2+2(m'_1-m_2)(m''_1+m_2) \Big\} \nonumber\\
&& +P^\m \ess \cdot P \Big\{ 4\Big(M'^2+M''^2-q^2 \nonumber\\
& & +2(m'_1-m_2)(m''_1+m_2) \Big)(A^{(2)}_2+A^{(2)}_3-A^{(1)}_1)
\nonumber\\
&&+2\Big( q^2-N'_1-N''_1-(m'_1+m''_1)^2 \Big)(A^{(1)}_1+A^{(1)}_2 -1) \Big\} \nonumber\\
&& + P^\m \frac{\ess \cdot \o}{\o \cdot P} \Big\{ 4\Big(M'^2+M''^2-q^2-2N_2+2(m'_1-m_2)(m''_1+m_2) \Big)B^{(2)}_1
\nonumber\\ 
&&+2\Big(q^2-N'_1-N''_1-(m'_1+m''_1)^2 \Big)C^{(1)}_1 \Big\} \nonumber\\
&& +q^\m \{\cdots\} + \o^\m \{ \cdots \} \, ,
\eea
where we have used the relation (A.7).

In the standard light-front formalism only  $\hat{A}^+$ and $\hat{B}^+$ are used
with the additional requirement that all quarks are on their respective mass shells, i.e.,
$N'_1$, $N''_1$ and $N_2$ are put equal to zero and $M'$, $M''$ are replaced by $M'_0$,
$M''_0$, respectively, on the right-hand-side of Eq.(C.1), but $P'^+ =P''^+$ is kept unchanged. 
In particular, 
$\hat{G}^+_0 =-2P'^+ A_0(q^2)$ is given by
\bea
A_0(q^2)&=&  \frac{N_c}{16\pi^3} \int ^1_0 dx \int d^2 p'_\perp
\frac{2h'_0 h''_0}{(1-x)\N'_1 \N''_1} \Bigg\{ 2xM''_0\Big[(1-x)m'_1+xm_2)\Big] \nonumber\\
&& +\frac{(2x-1)M''_0+m''_1-m_2}{(1-x)D''}\Bigg[p'_\perp p''_\perp \nonumber\\ 
& & +\Big(xm_2+(1-x)m'_1\Big)
\Big(xm_2-(1-x)m''_1 \Big) \Bigg] \Bigg\} \, ,
\eea
and the SLF expression for the form factor $f(q^2)$, denoted as  $f^{SLF}(q^2)$, is
\be
f^{SLF}(q^2)=-2M'' A_0(q^2)-(M'^2-M''^2-q^2) \; a_+(q^2) \, .
\ee

Yet another expression for $f(q^2)$ is found if the amplitude $\hat{G}^\m_+$ for on-shell
quarks is used, i.e., if the longitutinal mode is avoided. We denote it by $f^{CDKM}(q^2)$
since the CDKM approach of Ref. \cite{karmanov1},
gives the same result:
\bea
f^{CDKM}(q^2)&=&  \frac{N_c}{16\pi^3} \int ^1_0 dx \int d^2 p'_\perp
\frac{h'_0 h''_0}{(1-x)\N'_1 \N''_1} \Bigg\{-2m''_1 M'^2_0-2m'_1 M''^2_0+2m_2 q^2 \nonumber\\
&& +2(m'_1-m_2)(m''_1-m_2)(m'_1+m''_1)-8(m'_1-m_2)A^{(2)}_1 \nonumber\\
&&+\frac{4A^{(2)}_1}{D''} \Big[ M'^2_0+M''^2_0-q^2+2(m'_1-m_2)(m''_1+m_2) \Big] \Bigg\} \, .
\eea

Next we shall derive analoguous formulas for $f_V$.
Using the tensor decompositions of Appendix A the trace (3.13) can be
represented as
\bea
s^\m &=&\en^\m \Big\{ 8A^{(2)}_1+2 \Big(M^2-N'_1-N_2-(m_1-m_2)^2 \Big) \Big\} \nonumber\\
&&+P^\m \frac{\en \cdot \o}{\o \cdot P} \Big\{ 8B^{(2)}_1-2C^{(1)}_1 \Big\}
+\o^\m \frac{\en \cdot \o}{(\o \cdot P)^2}2C^{(2)}_2 \nonumber\\
&&-\frac{2}{D'} \Big\{ \en^\m \; 4 (m_1+m_2)A^{(2)}_1 \nonumber\\
&&+P^\m \frac{\en \cdot \o}{\o \cdot P}\Big( 4(m_1+m_2)B^{(2)}_1-2m_1 C^{(1)}_1 \Big)
+\o^\m \frac{\en \cdot \o}{(\o \cdot P)^2}(m_1+m_2)C^{(2)}_2 \Big\} \, . \nonumber\\
\eea
The standard light-front expression for the vector decay constant, $f^{SLF}_V$, can
be derived for on-shell quarks \cite{jaus4}:
\bea
f^{SLF}_V = \frac{N_c}{8\pi^3} \int ^1_0 dx \int d^2 p'_\perp
\frac{\sqrt{2} h'_0}{(1-x) \N'_1} \frac{M}{M'_0} \Bigg\{ xM'^2_0 -m'_1 (m'_1 -m_2 )
-p'^2_\perp  \nonumber\\
+ \frac{m'_1+m_2}{D'} p'^2_\perp  \Bigg\} \, .
\eea
A derivation based upon the amplitude $\hat{g}^{\perp}_+$, again for on-shell quarks, 
or that uses the approach of Ref. \cite{karmanov1} leads to
\bea
f^{CDKM}_V = \frac{N_c}{8\pi^3} \int ^1_0 dx \int d^2 p'_\perp
\frac{\sqrt{2} h'_0}{(1-x) \N'_1}  \Bigg\{ xM'^2_0 -m'_1 (m'_1 -m_2 )
-p'^2_\perp  \nonumber\\
+ \frac{m'_1+m_2}{D'} p'^2_\perp  \Bigg\} \, .
\eea

\end{appendix}

\def\etal{et al.}
\gdef\journal#1, #2, #3, #4 { {\sl #1~}{\bf #2}\ (#3)\ #4 }
\def\pr{\journal Phys. Rev., }
\def\prd{\journal Phys. Rev. D, }
\def\prl{\journal Phys. Rev. Lett., }
\def\jmp{\journal J. Math. Phys., }
\def\np{\journal Nucl. Phys., } 
\def\pl{\journal Phys. Lett., }

\end{document}